\documentclass[aps,prb,twocolumn,superscriptaddress,longbibliographys]{revtex4-2}
\usepackage{graphicx}
\usepackage{bm}% bold math
\usepackage{natbib}
\usepackage{color}
\usepackage{epstopdf}
\usepackage{amsmath} 
\usepackage{url}
\usepackage{braket}
\usepackage{mathtools}
\usepackage{xparse}
\definecolor{darkGreen}{RGB}{0,110,0}
\definecolor{darkBlue}{RGB}{0,0,130}
\usepackage[colorlinks=true, urlcolor=darkBlue,citecolor=darkGreen,linkcolor=darkBlue]{hyperref}
\begin{document}

%\preprint{}

\title{Squashed entanglement in one-dimensional quantum matter}


\author{Alfonso Maiellaro}
\affiliation{Dipartimento di Fisica "E.R. Caianiello", Università di Salerno, Via Giovanni Paolo II, 132, I-84084 Fisciano (SA), Italy}
\author{Francesco Romeo}
\affiliation{Dipartimento di Fisica "E.R. Caianiello", Università di Salerno, Via Giovanni Paolo II, 132, I-84084 Fisciano (SA), Italy}
\affiliation{INFN, Sezione di Napoli, Gruppo collegato di Salerno,Italy}
\author{Roberta Citro}
\affiliation{Dipartimento di Fisica "E.R. Caianiello", Università di Salerno, Via Giovanni Paolo II, 132, I-84084 Fisciano (SA), Italy}
\affiliation{INFN, Sezione di Napoli, Gruppo collegato di Salerno, Italy}
\author{Fabrizio Illuminati}
\email[Corresponding author: ]{filluminati@unisa.it}
\affiliation{Dipartimento di Ingegneria Industriale, Università di Salerno, Via Giovanni Paolo II, 132, I-84084 Fisciano (SA), Italy}
\affiliation{INFN, Sezione di Napoli, Gruppo collegato di Salerno,Italy}

%\thanks{}


\date{March 6, 2023}

\begin{abstract}
Squashed entanglement and its universal upper bound, the quantum conditional mutual information, are faithful measures of bipartite quantum correlations defined in terms of multipartitions. As such, they are sensitive to the fine-grain structure of quantum systems. Building on this observation, we introduce the concept of quantum conditional mutual information between the edges of quantum many-body systems. We show that this quantity characterizes unambiguously one-dimensional topological insulators and superconductors, being equal to Bell-state entanglement in the former and to half Bell-state entanglement in the latter, mirroring the different statistics of the edge modes in the two systems. The edge-to-edge quantum conditional mutual information is robust in the presence of disorder or local perturbations, converges exponentially with the system size to a quantized topological invariant, even in the presence of interactions, and vanishes in the trivial phase. We thus conjecture that it coincides with the edge-to-edge squashed entanglement in the entire ground-state phase diagram of symmetry-protected topological systems, and we provide some analytical evidence supporting the claim. By comparing them with the entanglement negativity, we collect further indications that the quantum conditional mutual information and the squashed entanglement provide a very accurate characterization of nonlocal correlation patterns in one-dimensional quantum matter.
\end{abstract}

%\keywords{}

\maketitle

\section{introduction}
Identifying entanglement-based order parameters able to characterize a large variety of quantum phases and at the same time to discriminate between different forms of quantum orders has remained a major challenge in condensed matter physics for the last two decades. The block von Neumann entanglement entropy \cite{PhysRevLett.78.2275,PhysRevLett.122.210402,PhysRevB.86.094412} and the block entanglement spectrum \cite{PhysRevA.78.032329,PhysRevLett.113.060501,PhysRevB.81.064439,PhysRevLett.105.080501} in simple bipartite systems have become central tools for the characterization of quantum collective behaviors, including topologically ordered phases \cite{Sato_2017,Maiellaro2019,Alicea_2012,MaiellaroEPJplus,doi:10.1146,PhysRevLett.96.110404,MaiellaroEPJst,PhysRevLett.105.177002,MaiellaroRomeoIlluminati2022,Maiellaro_2020,MaielFlux}. Indeed, nontrivial topological order in two-dimensional systems has been identified by means of the sub-leading term to the block von Neumann entanglement entropy, the so called topological entanglement entropy (TEE) \cite{HAMMA200522,PhysRevLett.96.110404,PhysRevLett.96.110405}. 

When considering either one-dimensional or higher-dimensional systems with bulk-edge correspondence, both the block von Neumann entanglement entropy and the block entanglement spectrum fail to discriminate between topologically ordered and standard Ginzburg-Landau ordered phases, associated with spontaneous symmetry breaking and nonvanishing order parameter. In fact, these measures based on simple bipartitions of a system into two parts (blocks) cannot account for the different physical properties of the bulk and of the edges between topologically trivial and topologically nontrivial phases of quantum matter \cite{PhysRevB.81.064439}. Moreover, for all open quantum systems in or out of equilibrium in any dimension, block entropies and entanglement spectra necessarily include contributions from both classical and quantum fluctuations and thus cease to be valid and meaningful measures of nonlocal quantum correlations.

Following some preliminary efforts to address the problem \cite{Wen2019,PhysRevB.101.085136,10.21468/SciPostPhysCore.3.2.012}, recently two well-defined multipartion-based measures of bipartite entanglement and bipartite correlations, respectively the squashed entanglement (SE) and the quantum conditional mutual information (QCMI), have been introduced as possible efficient tools in the study of one-dimensional quantum matter, given their ability to detect the fine-grain structures of quantum systems and thus, in particular, to discriminate between the different bulk and edge contributions to the long-distance correlation patterns in topologically trivial and topologically nontrivial quantum many-body systems  \cite{MaiellaroIlluminati2022}. These two quantities are intimately related, as the SE is defined as the infimum of the QCMI over all possible state extensions. Therefore, the QCMI is always an upper bound to the SE.
 
Specifically, for many-body systems with open boundary conditions we introduced the edge-to-edge QCMI, thus being an upper bound to the SE between the system edges, and we found that it defines the natural quantized, non-local order parameter for Kitaev topological superconductors in one spatial dimension and in quasi one-dimensional geometries \cite{MaiellaroIlluminati2022}. For such systems, the QCMI exhibits the correct scaling at the quantum phase transition, is stable in the presence of interactions and robust against the effects of disorder and local perturbations. We introduced two distinct multipartition-based forms of the QCMI: the tripartition-based edge-edge QCMI $I_{(3)}(A\!\!:\!\!B \, | \, C)$ corresponding to a edge $A$ - entire bulk $C$ - edge $B$ tripartition, which leads to a phase diagram equivalent to that of the corresponding spin chain obtained via a Jordan-Wigner transformation, and the quadripartition-based edge-edge QCMI 
$I_{(4)}(A\!\!:\!\!B \, | \, C_1)$, corresponding to a edge $A$ - bipartite bulk $C = C_1 C_2$  - edge $B$ quadripartition and a partially traced-out bulk (e.g. by tracing out $C_2$), which discriminates between symmetric topological regimes and ordered phases with spontaneously broken symmetries \cite{Franchini2017,condmat6020015,condmat7010026}.

Motivated by these results, in the present work we provide an in-depth study of the edge-to-edge QCMI and SE and show that they characterize topological quantum phase transitions in 
one-dimensional systems. Specifically, we investigate and compare the two paradigmatic models describing one-dimensional topological insulators and superconductors, respectively the Su-Schrieffer-Heeger (SSH) insulating chain \cite{PhysRevLett.42.1698} and the Kitaev superconducting wire \cite{Kitaev_2001}. For these systems we show that the edge-to-edge bipartite QCMI identifies the correct nonlocal order parameter that singles out the phase transitions, characterizes the topologically ordered phases of topological insulators and superconductors, and discriminates between them. 

The edge-to-edge QCMI is robust under variations of the sample conditions due to disorder or local perturbations, and scales exponentially with the size of the system, converging to a quantized, topologically invariant value even in the presence of interactions. Crucially, the QCMI is sensitive to the different nature of the edge modes. Indeed, it takes different quantized values, respectively to Bell-state entanglement and half Bell-state entanglement, depending on the statistics of the topological edge modes, respectively Dirac fermions in the SSH chain and Majorana fermions ("half-Dirac fermions") in the Kitaev wire. 

Having found that the edge-to-edge QCMI in low-dimensional topological systems exhibits the same behaviour expected for the genuine SE between the system edges, we conjecture that the two quantities do actually coincide, and for small-size systems we provide further supporting analytical evidence to this statement. In general, while the computational effort in the evaluation of the SE is strongly dependent on the system size and generically a $NP$-hard problem, evaluating the QCMI requires only a limited amount of computational resources. This feature of the QCMI between the edges of a topological system is due to the fact that in the latter the bulk does not contribute to the quantum correlations between the edges. Therefore the QCMI is insensitive to different partitions of the bulk and to the system size as soon as the latter exceeds a (small) critical threshold value.

We also compare the edge-to-edge QCMI with the entanglement negativity, a measure of bipartite entanglement that is widely used in the study of quantum statistical mechanics and quantum matter because of its simplicity and computability \cite{PhysRevA.65.032314}. We identify some possible failures of the latter and provide further evidence that multipartition-based measures such as the edge-to-edge SE and QCMI are indeed the natural framework for the characterization of one-dimensional quantum matter. Finally, we discuss the perspectives for the experimental accessibility of SE/QCMI as well as their generalization to many-body systems in higher spatial dimensions.

The paper is organized as follows. In Sec. \ref{SecII} we introduce the QCMI, the SE and the two fundamental forms of upper bounds of SE based on the QCMI, $I_{3}$ and $I_{4}$. We also review the main properties of the one-dimensional SSH and Kitaev models. In Sec. \ref{SecIII} we study the main features of the QCMI in the two systems, including the effects of interactions, and compare the behavior of the QCMI with that of the entanglement negativity. In Sec. \ref{SecIV} we investigate and discuss the conjectured equivalence between the QCMI and the SE. In Sec. \ref{SecV} we discuss our results, consider possible experimental protocols to measure the QCMI and the SE, and discuss generalizations and further applications in the study of two-dimensional quantum matter. Technical details and mathematical methods used troughout the paper are reported in the Appendices. Finite-size scaling at phase transition boundaries and robustness against disorder of the QCMI in the topological phases are reported respectively in Appendices \ref{AppA} and \ref{AppB}, while in Appendix \ref{AppC} we analyze the Jordan-Wigner transformation applied to the two systems and discuss the phase diagrams of the interacting SSH model.

\section{Theory}
\label{SecII}
Consider a quantum system $G$, two arbitrary subsystems $A$ and $B$, and a reminder $C$, such that $ABC$ defines a tripartition of $G$. Consider next a partition of the reminder: $C=C_1C_2$, so that now $ABC_1C_2$ defines a quadripartition of $G$. One can then introduce two inequivalent measures 
$I_{(3)} = I(A\!\!:\!\!B \, | \, C)$ and $I_{(4)} = I(A\!\!:\!\!B \, | \, C_1)$ of the QCMI between subsystems $A$ and $B$, respectively conditioned to $C$ and to $C_1$ once $C_2$ has been traced out \cite{MaiellaroIlluminati2022}:
\begin{equation}
\label{upper3}
I_{(3)} = \frac{1}{2} \biggl[ S(\rho_{AC}) + S(\rho_{BC}) - S(\rho_{ABC}) - S(\rho_{C})\biggr] \, , 
\end{equation}
and
\begin{equation}
\label{upper4}
I_{(4)} = \frac{1}{2} \biggl[S(\rho_{AC_1}) + S(\rho_{BC_1}) - S(\rho_{ABC_1}) - S(\rho_{C_1})\biggr] \, ,
\end{equation}
where $S(\rho)$ is the von Neumann entropy of the quantum state $\rho$. The expressions $I_{(3)}$ and $I_{(4)}$ are the only two inequivalent QCMIs corresponding to the same reduced state $\rho_{AB}$ obtained from the global state $\rho_{G}$ \cite{MaiellaroIlluminati2022}. On the other hand, the SE ($E_{sq}(\rho_{XY})$) \cite{Tucci2002,Christandl2004} of a bipartite system $X \cup Y$ is defined in terms of the QCMI as:
\begin{eqnarray}
E_{sq}(\rho_{XY})=\inf_{\rho_{XYZ}} \bigg\{ I(X:Y|Z) \bigg\},
\label{SE}
\end{eqnarray}
where $\inf_{\rho_{XYZ}} \bigg\{ I(X:Y|Z) \bigg\}$ denotes the infimum of $I(X:Y|Z)$ over all quantum states $\rho_{XYZ}$ of all possible extensions $X \cup Y \cup Z$, with fixed subsystems $X$ and $Y$. Given a state $\rho_G$ of the entire system $G$, then $\rho_{AC}=Tr_B(\rho_{G})$,  $\rho_{BC}=Tr_A(\rho_{G})$,  $\rho_{C}=Tr_{AB}(\rho_{G})$,  $\rho_{AC_1}=Tr_{BC_2}(\rho_{G})$,  $\rho_{BC_1}=Tr_{AC_2}(\rho_{G})$,  $\rho_{ABC_1}=Tr_{C_2}(\rho_{G})$ and $\rho_{C_1}=Tr_{ABC_2}(\rho_{G})$ are the reduced states of the subsystems involved. 
SE is the "perfect" measure of entanglement on all quantum states as it is the only entanglement monotone that is also convex, additive, asymptotically continuous, faithful, monogamous, and reduces to the von Neumann entanglement entropy on pure states, thus satisfying all the required properties for a \emph{bona fide} entanglement measure \cite{Bengtsson2020,Horodecki2009,MaiellaroIlluminati2022}. 

In complete generality, it trivially holds that $I_{(3)} \geq E_{sq}(\rho_{AB})$ and 
$I_{(4)} \geq E_{sq}(\rho_{AB})$. On the other hand, if $A$ and $B$ are the edges of a one-dimensional many-body system with open boundary conditions, we will find that a further chain inequality appears to hold: $I_{(3)} \geq I_{(4)} \geq E_{sq}(\rho_{AB})$. Actually, as discussed in Sec. \ref{SecIV} below, one can show numerically that equality holds at least for symmetry-protected topological systems of limited size: $I_{(3)}=I_{(4)}=E_{sq}(\rho_{AB})$. In Sec. \ref{SecIV} we also propose further arguments supporting the conjecture that the equality $I_{(3)}=I_{(4)}=E_{sq}(\rho_{AB})$ between the edge-to-edge QCMIs and the edge-to-edge SE holds for all one-dimensional symmetry-protected topological systems of arbitrary size. We will thus discuss the topological properties of the systems under investigation by considering $I_{(3)}$ and $I_{(4)}$ whenever they coincide and whenever they are expected to be equal to the exact SE $E_{sq}$.
\begin{figure}
	\includegraphics[scale=0.095]{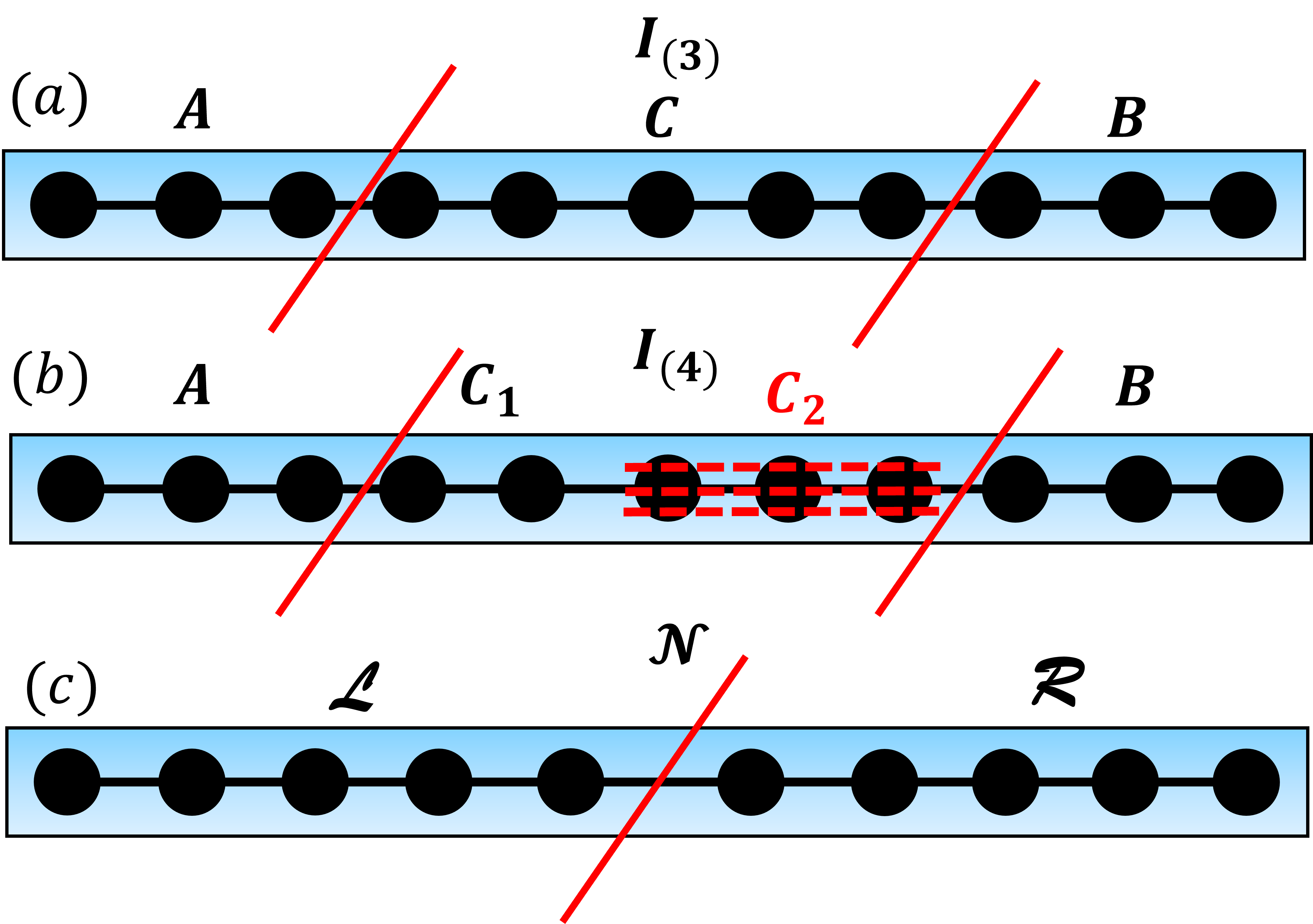}\\
	\vspace{0.5cm}
	\includegraphics[scale=0.13]{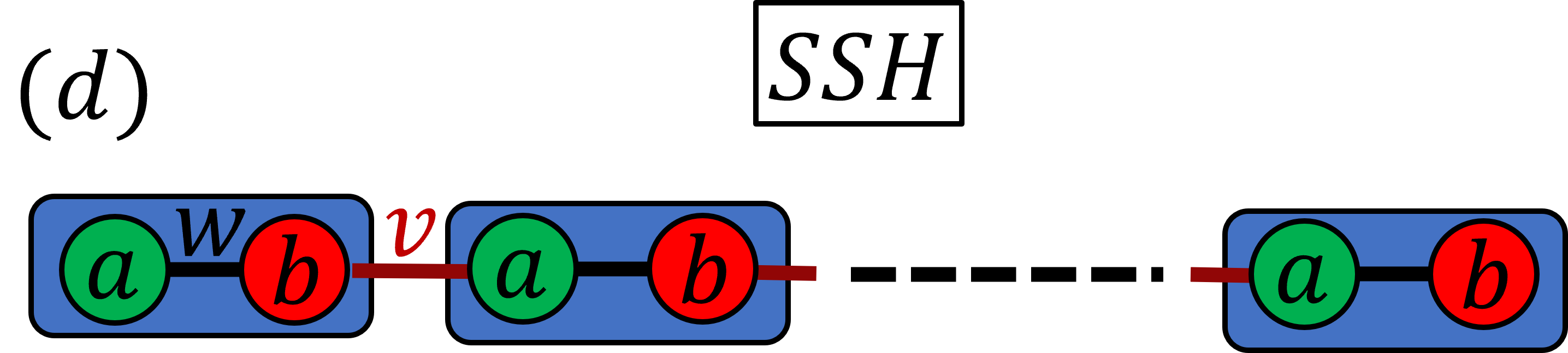}
	\hspace{1cm}
	\includegraphics[scale=0.12]{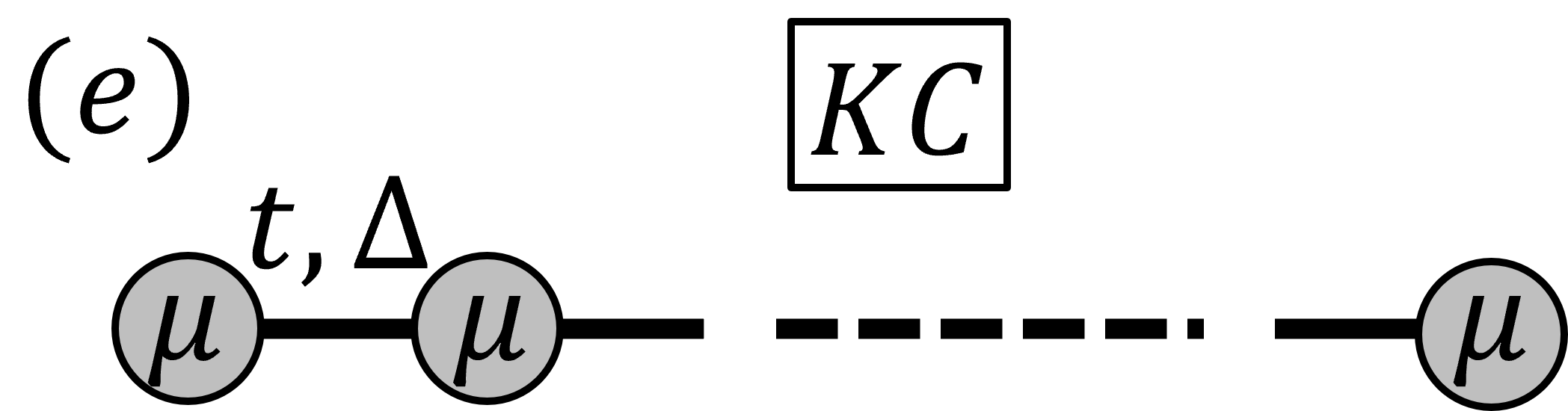}	
	\caption{Panels (a) and (b): system multipartitions associated to the definition of the corresponding bipartite, edge-to-edge SE and QCMIs. Panel (a): tripartition of a quantum chain into two edges $A$ and $B$ and a bulk $C$. Panel (b): quadripartition of the same chain, with a bipartite bulk $C=C_1C_2$ and part $C_2$ of the bulk traced out. Panel (c): standard bipartition of the same chain in two blocks (halves) for the study of the block entanglement negativity. Panels (d) and (e): sketch of the tight-binding models of the SSH and of the Kitaev chains. In panel (d) $a$ and $b$ are the two fermionic on-site species in the unit cell of the SSH model Hamiltonian.}
	\label{Figure1}
\end{figure}

In Fig. \ref{Figure1}, (a) and (b), we provide a sketch of the above multipartitions for a quantum system on a one-dimensional lattice. Subsystems $A$ and $B$ identify the system edges; $C$ identifies the full bulk. When the bulk is bipartite, i.e. $C=C_1C_2$, we denote by $C_1$ the part of the bulk that remains after part $C_2$ is traced out. Without loss of generality one can set the size of $C_1$ as $L_{C_1}=1$ and identify it with the site adjacent to edge $A$. We also fix the length of the edge partitions to $\lfloor L/3 \rfloor$, which is, for the system sizes considered, of the order of the decay length of the edge modes in the bulk.

In the present work we study and compare the SSH insulator and the Kitaev superconductor according to the multipartitions shown in Fig. \ref{Figure1} using the edge-to-edge QCMIs $I_{(3)}$ and $I_{(4)}$ to characterize the phase diagram and the topologically ordered phases. Moreover, we will consider a further estimator of bipartite entanglement, the so-called entanglement negativity ($\mathcal{N}$) \cite{PhysRevA.65.032314}, that essentially quantifies the Peres criterion for separability \cite{PhysRevLett.77.1413}. The negativity is overwhelmingly used in quantum information as well as in the study of quantum matter \cite{PhysRevLett.125.116801,PhysRevA.88.042318} for its simplicity and computability, notwithstanding that it fails to satisfy some of the most important properties required for a valid entanglement measure. Given a bipartition $\cal{L}\cal{R}$ into two adjacent blocks (halves) $\cal{L}$ and $\cal{R}$, as shown in Fig. \ref{Figure1} (c), the negativity $\mathcal{N}$ is defined as the trace norm of the partial transpose 
$\rho^{T_{\cal{L}}}$ of the bipartite state $\rho (\cal{L}\cal{R})$ with respect to one of the two blocks:
\begin{equation}
\mathcal{N}=\frac{||\rho^{T_{\cal{L}}}||_1-1}{2}=\sum_{i} |\lambda_i| \, ,
	\label{negativity}
\end{equation}
where $||\rho^{T_{\cal{L}}}||_1$ is the trace norm of $\rho^{T_{\cal{L}}}$ and 
the $\lambda_i$s are the negative eigenvalues of $\rho^{T_{\cal{L}}}$. Next, we recall the SSH and Kitaev model Hamiltonians:
\begin{eqnarray}
H_{SSH}=w \sum_{j=1}^{L} c^\dagger_{a,j} c_{b,j}+v \sum_{j=1}^{L-1} c^\dagger_{a,j+1}  c_{b,j}+h.c. \, ,
\label{SSHhamiltonian}
\end{eqnarray}
\begin{eqnarray}
	\begin{split}
H_K=&\sum_{j=1}^{L-1}\biggl(\Delta c^\dagger_{j+1}c^\dagger_{j}-tc^\dagger_{j}c_{j+1}+h.c.\biggr)+\\
&-\sum_{j=1}^L \mu \biggl(c^\dagger_{j}c_{j} -\frac{1}{2}\biggr) \, .
\end{split}
\label{Kitaev}
\end{eqnarray} 

The SSH chain, Eq. \ref{SSHhamiltonian}, describes spinless fermions with staggered hopping amplitudes $w$ and $v$. The two fermionic species $a$ and $b$ define the two different degrees of freedom per unit cell. The Kiatev chain, Eq. \ref{Kitaev}, describes spinless fermions with a $p$-wave superconducting pairing potential $\Delta$, an hopping strength $t$ and an on-site chemical potential $\mu$. 

These two models embed all the key properties of topological insulators and superconductors: an insulating bulk with boundary conduction, a protecting symmetry (chiral symmetry for the SSH model; particle-hole symmetry for the Kitaev model) and a bulk-edge correspondence. The topologically ordered phase occurs, respectively, when $w < v$ and when $\mu < 2t$, 
$\Delta \neq 0$. The crucial discriminant between the two models is that they belong to two distinct classes of the ten-fold classification \cite{Altland1997}, featuring topological edge modes of different physical nature: fermionic for the SSH insulator and Majorana for the Kitaev superconductor. When interactions are included, the additional interaction terms read:
\begin{eqnarray}
	\label{IntSSH}
	&&H_{I,SSH}=\sum_{j=1}^L \left[ U_1 n_{a,j} n_{b,j}+ U_2 n_{b,j} n_{a,j+1} \right] \, , \\
	\label{IntKitaev}
	&&H_{I,K}=\sum_{j=1}^L U (2 n_j-1)(2 n_{j+1}-1) \, .
\end{eqnarray}
where $n_j= c^\dagger_j c_j$. Unless otherwise stated, from now on we set $U_1=U_2=U$. 

\section{Results}
\label{SecIII}
\begin{figure}
	\includegraphics[scale=0.125]{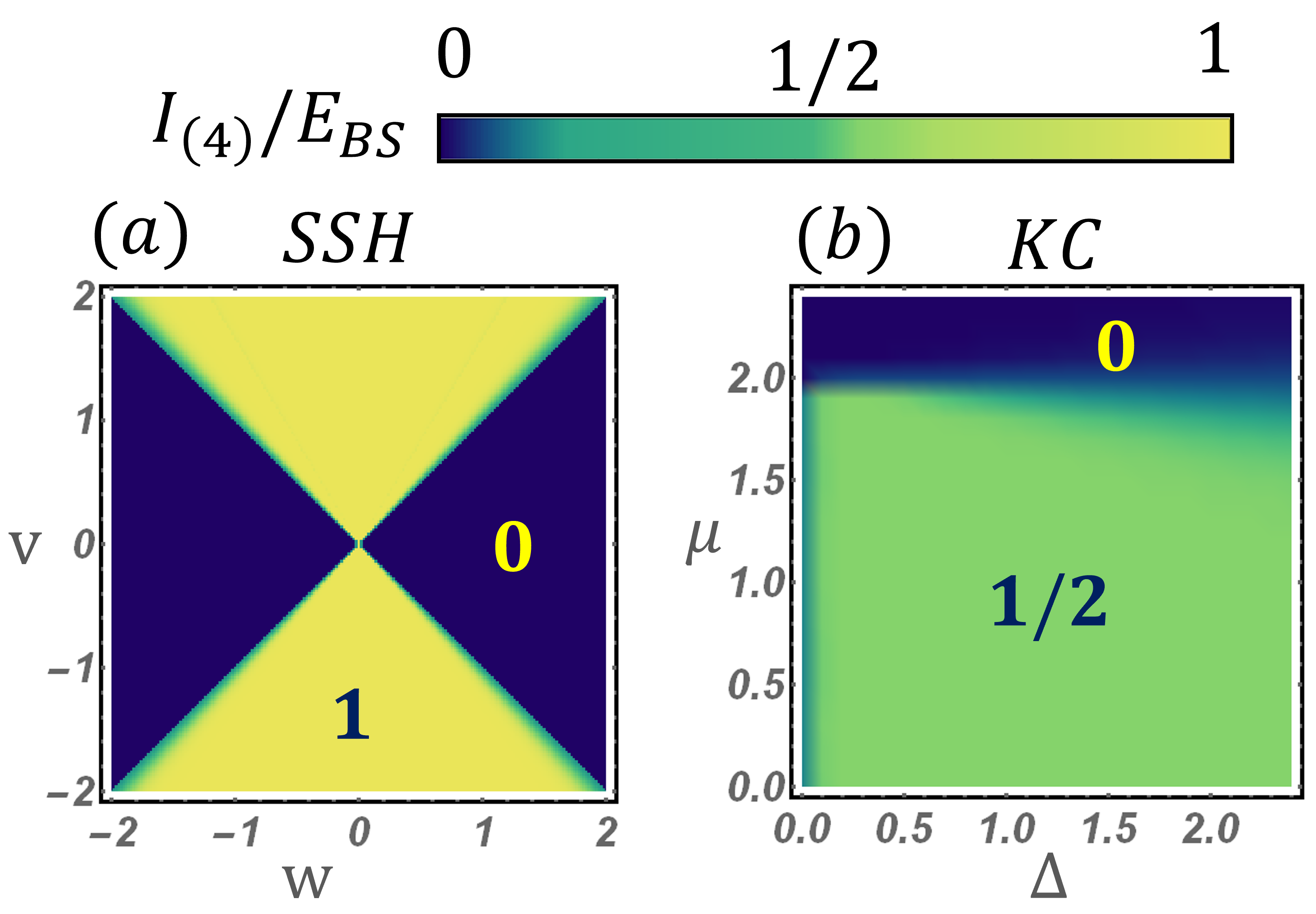}
	\includegraphics[scale=0.08]{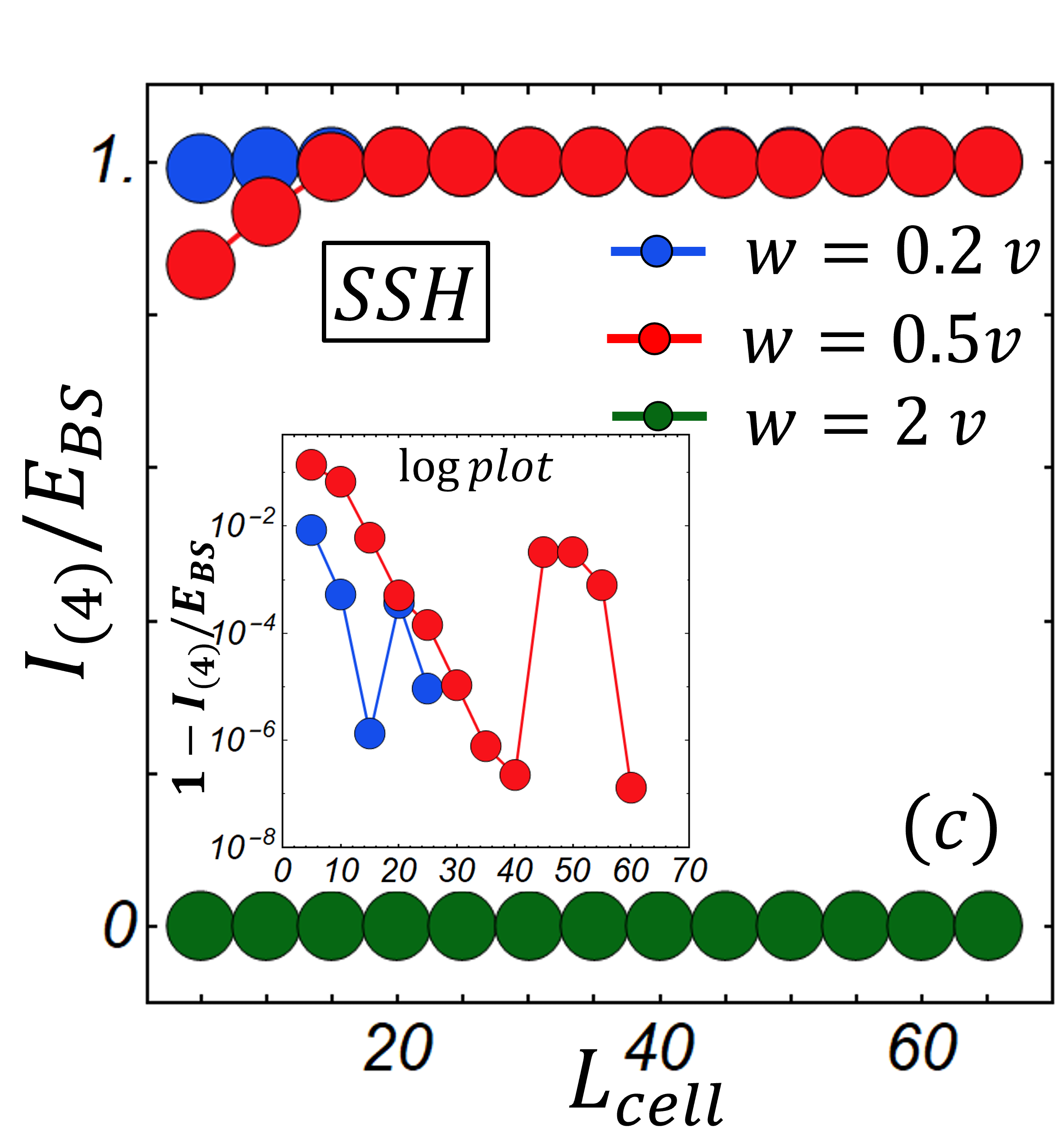}
	\includegraphics[scale=0.08]{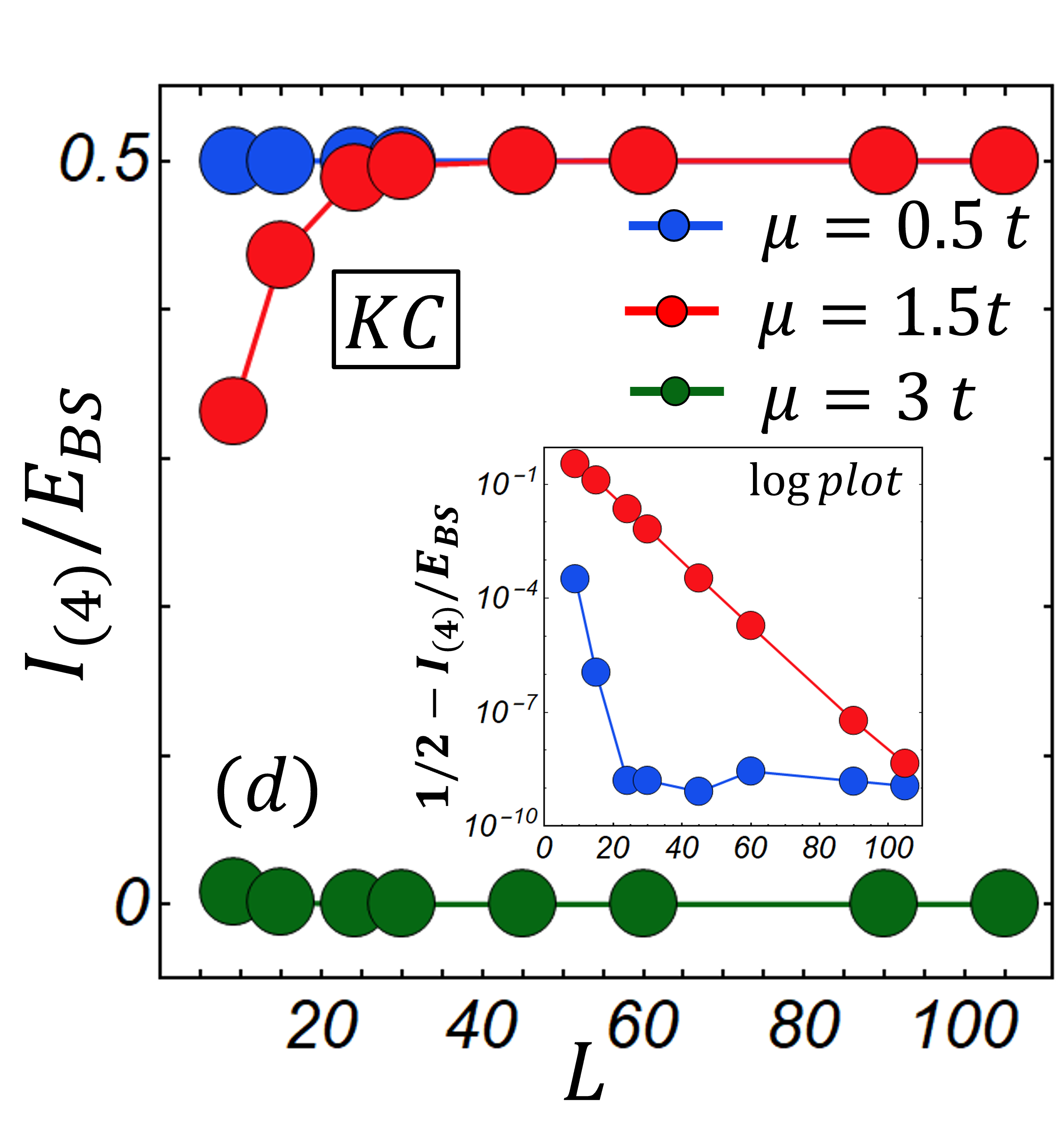}
	\caption{Phase diagrams as determined by the QCMI to Bell-state entanglement ratio $I_{(4)}/E_{BS}$. Panel (a): SSH chain of size $L_{cell}=50$ unit cells. Panel (b): Kitaev chain of size $L=60$ sites at $t=1$. Panels (c) and (d): scaling in the topological and in the trivial regimes, with $t=\Delta=1$ in panel (d). Insets: log-scale plots showing for both models the exponential convergence to the quantized ground-state value of the edge-to-edge SE $E_{sq}$.}
	\label{Figure2}
\end{figure}

In Fig. \ref{Figure2} (a) and (b), we report the phase diagrams of the two models determined by the ratio between the edge-to-edge QCMI $I_{(4)}$ and the reference Bell-state entanglement $E_{BS}=\ln 2$. The sizes of the two chains are identified by the number  of unit cells $L_{cell}$ for the SSH insulator and by the number of fermionic sites $L$ for the Kitaev superconductor. Throughout the entire topological phase $I_{(4)}/E_{BS}=1$ for the SSH chain and $I_{(4)}/E_{BS}=1/2$ for the Kitaev chain, reflecting the different statistics of the topological modes in the two models \cite{Pachos2012,Maiellaro2018,MaiellaroProc1}.
The conjectured coincidence of the SE with the QCMI implies that $E_{sq} = E_{BS}$ for the SSH insulator, identifying the presence of two fermionic edge modes, while for the Kitaev superconductor $E_{sq} = E_{BS}/2$, detecting the presence of two half-fermion Majorana edge modes. Finite-size effects are clearly visible near the phase boundaries $|w|=|v|$ and $|\mu|=2 t$. 

From Fig. \ref{Figure2} we see that in the topological phase the QCMI $I_{(4)}$ scales exponentially to the quantized values $E_{BS}$ and $E_{BS}/2$, while it remains pinned to zero in the trivial phase; since the SE is positive semi-definite and bounded from above by the QCMI, it follows that throughout the entire trivial phase it is certainly $I_{(4)} = E_{sq} = 0$. On the other hand, at the points of exact ground-state topological degeneracy, respectively $w=0$, $v \neq0$ and $|\mu|=0$, $t=\Delta$, both the topological fermionic modes and the Majoranas decouple from the bulk and nucleate at the edge of the chains. Correspondingly, the exact quantized value of the QCMI $I_{(4)}$ becomes independent of the chain size. 

It is important to observe that the same results are obtained resorting to the QCMI $I_{(3)}$. Indeed, due to the exponential decoupling of the edges from the insulating bulk, the definition of the edge-to-edge quantum conditional mutual information in a topological system is insensitive to the different partitions of the bulk, and therefore $I_{(3)} = I_{(4)}$ throughout the entire phase diagram. In fact, at the exact points of topological degeneracy, the two equal upper bounds $I_{(3)} = I_{(4)}$ on the edge-to-edge SE coincide with the quantized values $E_{BS}$ and $E_{BS}/2$. Away from the points of exact ground-state degeneracy but still inside the topologically ordered phases, numerical analysis confirms that, for small-sized systems, $I_{(3)}$ and $I_{(4)}$ still coincide and remain constant and equal, respectively, to the Bell-state entanglement $E_{BS}$ for the SSH insulator and to half the Bell-state entanglement $E_{BS}/2$ for the Kitaev superconductor. These findings, together with the property of asymptotic continuity enjoyed by SE strongly suggest that $I_{(3)}$ and $I_{(4)}$ coincide with the true SE $E_{sq}(\rho_{AB})$ between the edges not only in the trivial phase but also throughout the entire topological phase. Therefore, the edge-to-edge SE and, equivalently, the two edge-to-edge QCMIs $I_{(3)}$ and $I_{(4)}$ all appear to identify the same correct non-local order parameter for one-dimensional topological insulators and superconductors. 

In conclusion, all of the above leads us to conjecture that the edge-to-edge SE coincides exactly with the edge-to-edge QCMIs throughout the entire ground-state quantum phase diagrams of topological insulators and superconductors. We discuss in more detail the conjecture in Sec. \ref{SecIV} below, where we provide analytical and numerical arguments that: i) the QMCIs coincide with the true edge-to-edge SE in small systems; ii) the QMCIs scale exponentially to the SE shared by the non-trivial edge modes in the topologically ordered phases; iii) such SE between the system edges depends on the statistics of the edge modes; iv) the QCMIs are robust against disorder and local perturbations. Details on finite-size scaling at the phase transition boundaries are provided in Appendix \ref{AppA}, while point iv) is discussed at length in Appendix \ref{AppB}.

\begin{figure}
	\includegraphics[scale=0.125]{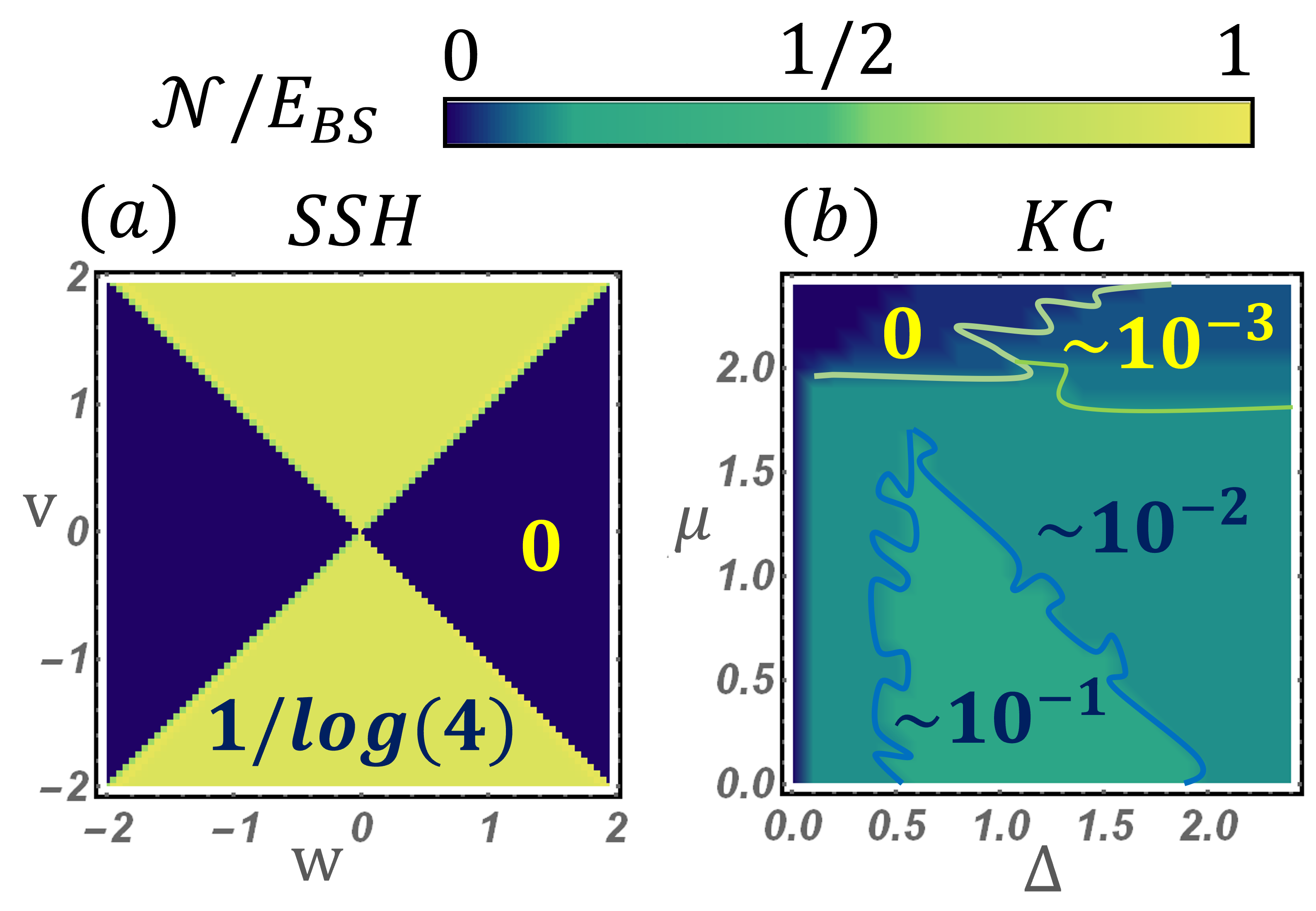}
	\includegraphics[scale=0.08]{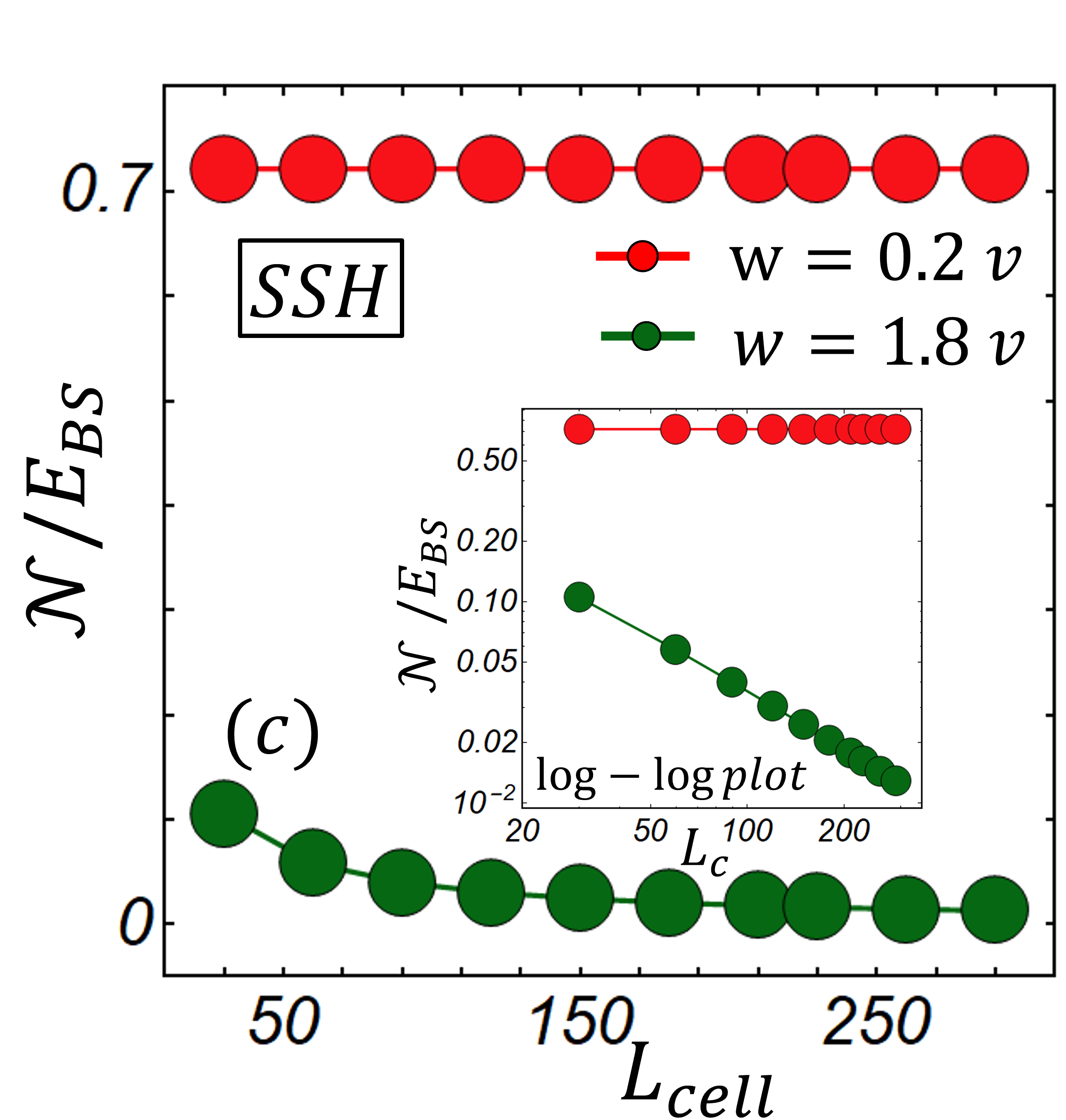}
	\includegraphics[scale=0.08]{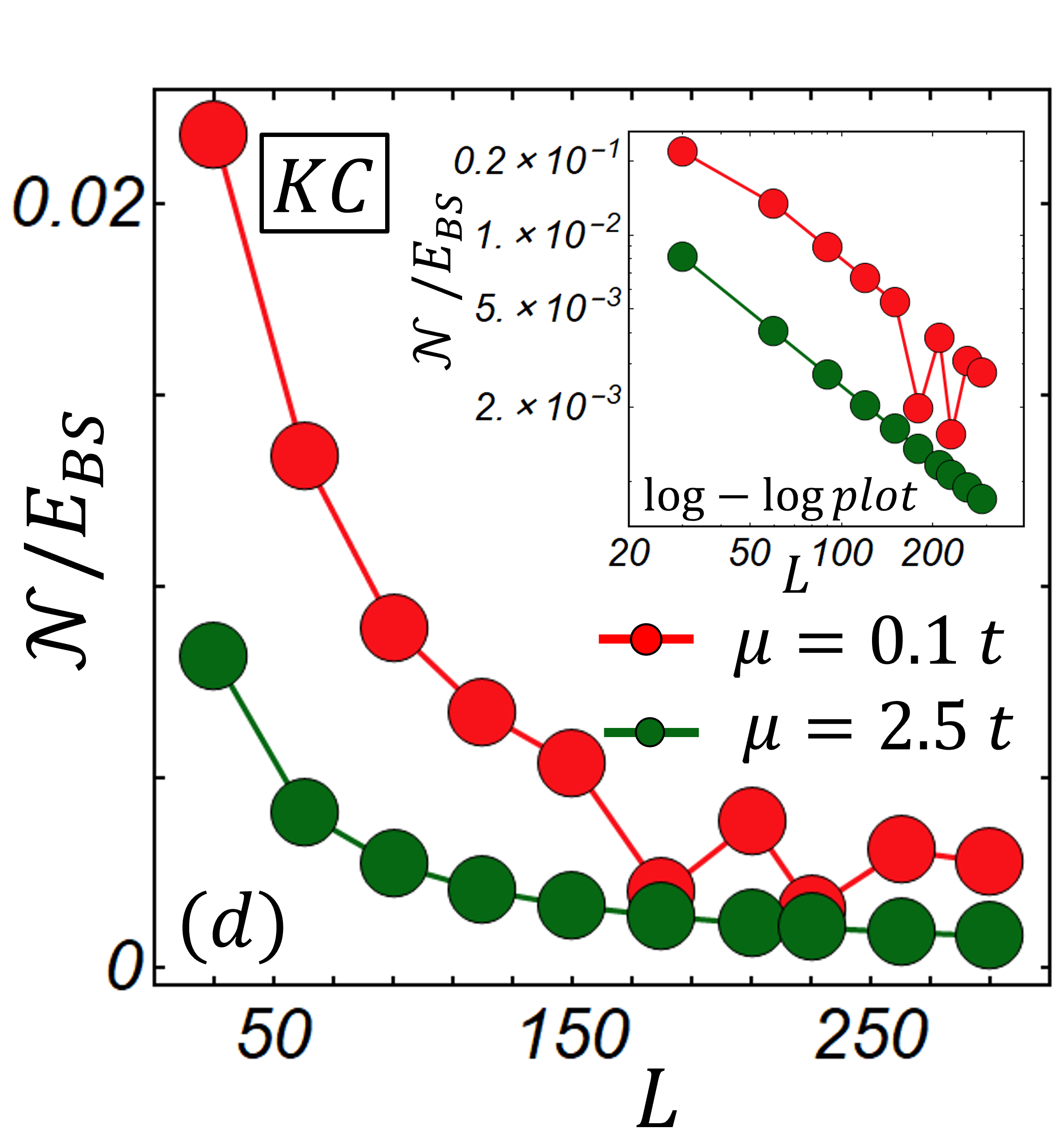}
	\caption{Phase diagrams as determined by the entanglement negativity to Bell-state entanglement ratio $\mathcal{N}/E_{BS}$. Panel (a): SSH chain of size $L_{cell}=50$ unit cells. Panel (b): Kitaev chain of size $L=60$ sites at $t=1$. Panels (c) and (d): scaling in the topological and in the trivial regimes, with $t=1$ and $\Delta=1.8$ in panel (d).}
	\label{Figure3}
\end{figure} 

As already mentioned, the entanglement negativity has become an increasingly popular tool in the investigation of topological quantum matter \cite{PhysRevA.65.032314,PhysRevLett.125.116801,PhysRevA.88.042318,Zimboras2015,Eisert2018}. Here we show that in fact at least some forms of
$\mathcal{N}$ do not provide the correct characterization of topological superconductors. In Fig. \ref{Figure3} we report the phase diagrams of the SSH and Kitaev chains as determined by the ratio $\mathcal{N}/E_{BS}$. For the SSH insulator we have that 
$\mathcal{N}=E_{BS}/\log 4$ in the nontrivial phase and $\mathcal{N}=0$
otherwise. On the other hand, $\mathcal{N}$ fails to reproduce the correct phase diagram of the Kitaev superconductor: from Fig. \ref{Figure3} we see that $\mathcal{N}$ vanishes asymptotically with increasing size of the system. This feature is a manifestation of the possible unfaithfulness of the negativity. On the contrary, the SE $E_{sq}$ is faithful, and thus $E_{sq}=0$ is a necessary and sufficient condition for separability. The above analysis confirms that, at variance with bipartition-based estimators of bipartite entanglement, the paradigm of multipartition-based bipartite SE appears to represent the correct framework for the investigation of one-dimensional quantum matter. It remains to be seen whether a more careful and sophisticated approach to the definition of the entanglement negativity in fermionic systems, such as that pioneered by Shinsei Ryu and coworkers \cite{Ryu2017,Ryu2019}, can lead to a characterization of topological systems that not only detects the transition, but also features the correct invariant topological quantization in the ordered phase, discriminates among different topological systems and edge modes, and discriminates between symmetric topological order and ordered phases associated to spontaneous symmetry breaking.

\begin{figure}
	\includegraphics[scale=0.12]{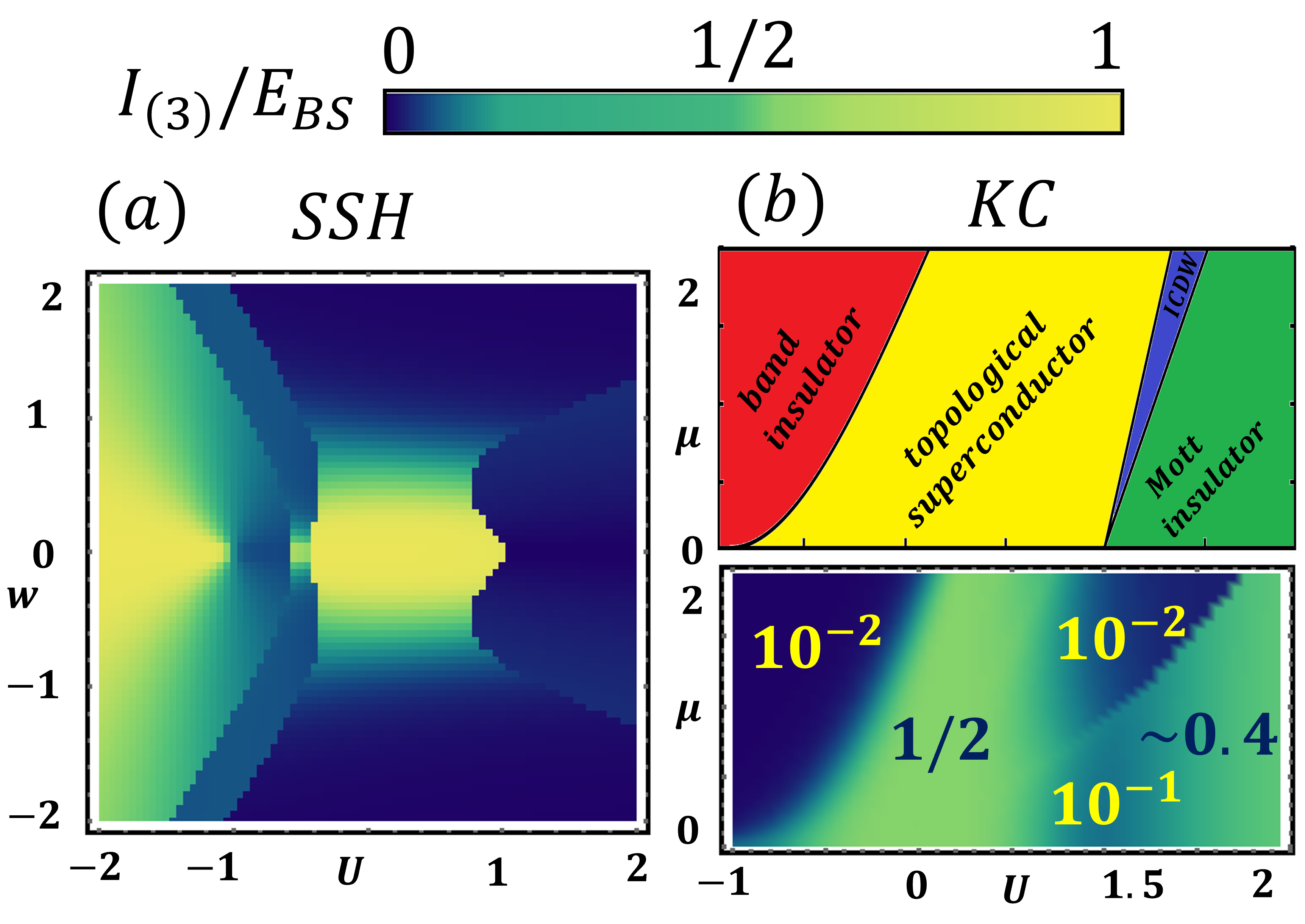}
	\caption{Phase diagrams of the interacting SSH and Kitaev chains determined by the QCMI to Bell-state entanglement ratio $I_{(3)}/E_{BS}$ as a function of the interaction strength $U$. In panel (a) the SSH chain size is $L_{cell}=14$ and $v=1$. In panel (b) the length of the Kitaev chain is $L=18$, while $t=\Delta=1$. The drawn-in phase diagram of the interacting Kitaev chain is taken from Ref. \cite{PhysRevB.92.115137}.}
	\label{Figure4}
\end{figure}

When interactions are included, it is convenient to map fermionic systems into interacting spin chains via the Jordan-Wigner mapping \cite{Franchini2017} (see Appendix \ref{AppC} for details). Although this procedure extends the Hilbert space dimension from $2 L$ to $2^L$, resorting to the QCMI $I_{(3)}$ allows to investigate the effects of interactions without numerical approximations. Therefore, for small system size, one can perform a direct comparison between well-known results coming from exact diagonalization and from numerical techniques \cite{10.21468/SciPostPhysLectNotes.8,ORUS2014117}. In Fig. \ref{Figure4} we report the contour plots of the QCMI to Bell-state entanglement ratio $I_{(3)}/E_{BS}$ for the SSH and Kitaev model Hamiltonians with the addition of the interaction terms $H_{I,SSH}$ and $H_{I,KC}$. 
From Fig. \ref{Figure4} (a) we see that in the presence of repulsive interactions the mean field topologically ordered phase $|w| \leq |v|$ is progressively reduced, until a trivial phase is reached independently of $w$ for $U > 1$. On the other hand, when we consider attractive interactions, a new region with $I_{(3)}/E_{BS}=1$ progressively reopens, being independent of $w$ for strong enough attraction ($U < -1$). The detailed analysis is illustrated in Appendix \ref{AppC}. 

In Fig. \ref{Figure4} (b) we compare the phase diagram of the interacting Kitaev chain determined by the ratio $I_{(3)}/E_{BS}$ with the standard reference one obtained by a variety of analytic and numerical methods in the recent literature \cite{PhysRevB.92.115137}. We find an excellent qualitative and quantitative agreement between both diagrams when the system behaves as a band insulator and a topological superconductor. Indeed, both diagrams match the same band insulator-topological superconductor phase boundary. $I_{(3)}$ also signals the presence of an incommensurate charge density wave phase (ICDW), in fair agreement with the most recent DRMG results \cite{Miao,PhysRevB.101.085125}. 

The phase diagram of the interacting Kitaev chain is currently understood by means of combination of several numerical techniques, like DMRG and bosonization \cite{ORUS2014117,10.21468/SciPostPhysLectNotes.8,Solr-2724221,PhysRevB.82.235120,PhysRevB.62.7019}, that involve various adjustments on the ground state parity, oscillatory behaviors of the ground-state wave functions and interpolation schemes on few numerical points \cite{Hassler_2012,Miao,PhysRevB.101.085125}, so that discrepancies and uncertainties emerge, even concerning a possible additional phase with odd ground-state parity between the ICDW and the Mott insulator phases \cite{PhysRevB.101.085125}, so that a definite phase boundary for the ICDW phase has yet to be identified. 

In the $I_{(3)}$-based phase diagram, the QCMI $I_{(3)}=0$, and therefore $E_{sq} = 0$ both in the trivial insulator and in the ICDW phases, signaling the absence of long-distance entanglement between the edges. On the other hand, $I_{(3)}/E_{BS}  \sim 1/2$ in the Mott insulator phase, implying a nontrivial long-distance edge-to-edge squashed entanglement $E_{sq}$, provided the conjectured coincidence between QCMI and SE holds. This finding yields strong support to recent studies suggesting the presence of topological order in the Mott regime of one-band fermionic systems \cite{PhysRevB.102.081110}. In this picture, the metal-insulator transition is equivalent to a topological transition via a mid-gap pole in the self-energy that matches the spectral pole of the localized surface state in a topological insulator. The QCMI and SE paradigm encapsulates such property of the Mott insulating phase by yielding a nontrivial quantized value of the QCMI to Bell-state entanglement ratio $I_{(3)}/E_{BS}$. In this respect, the difference between the numerical evaluation and the expected value $1/2$ for the ratio $I_{3}/E_{BS}$ in the Mott phase could originate from residual finite-size effects, obeying a slower finite-size scaling compared to the other phases.

\section{Equivalence between QCMI and SE: some approximate analytical results}
\label{SecIV}

The SE ($E_{sq}(\rho_{XY})$) introduced in Eq. \ref{SE} measures the bipartite entanglement between subsystems $X$ and $Y$ in the joint quantum state $\rho_{XY}$. It is defined in terms of nonoverlapping distinguishable multipartitions, for any spatial dimension, at any temperature, and on all quantum states (pure or mixed). The SE is defined as the infimum of the QCMI, taken over all the quantum-state extensions of arbitrary size $\rho_{XYZ}$ such that $\rho_{XY} = Tr_{Z}(\rho_{XYZ})$.

\begin{figure*}
	\label{AppA}
	\includegraphics[scale=0.12]{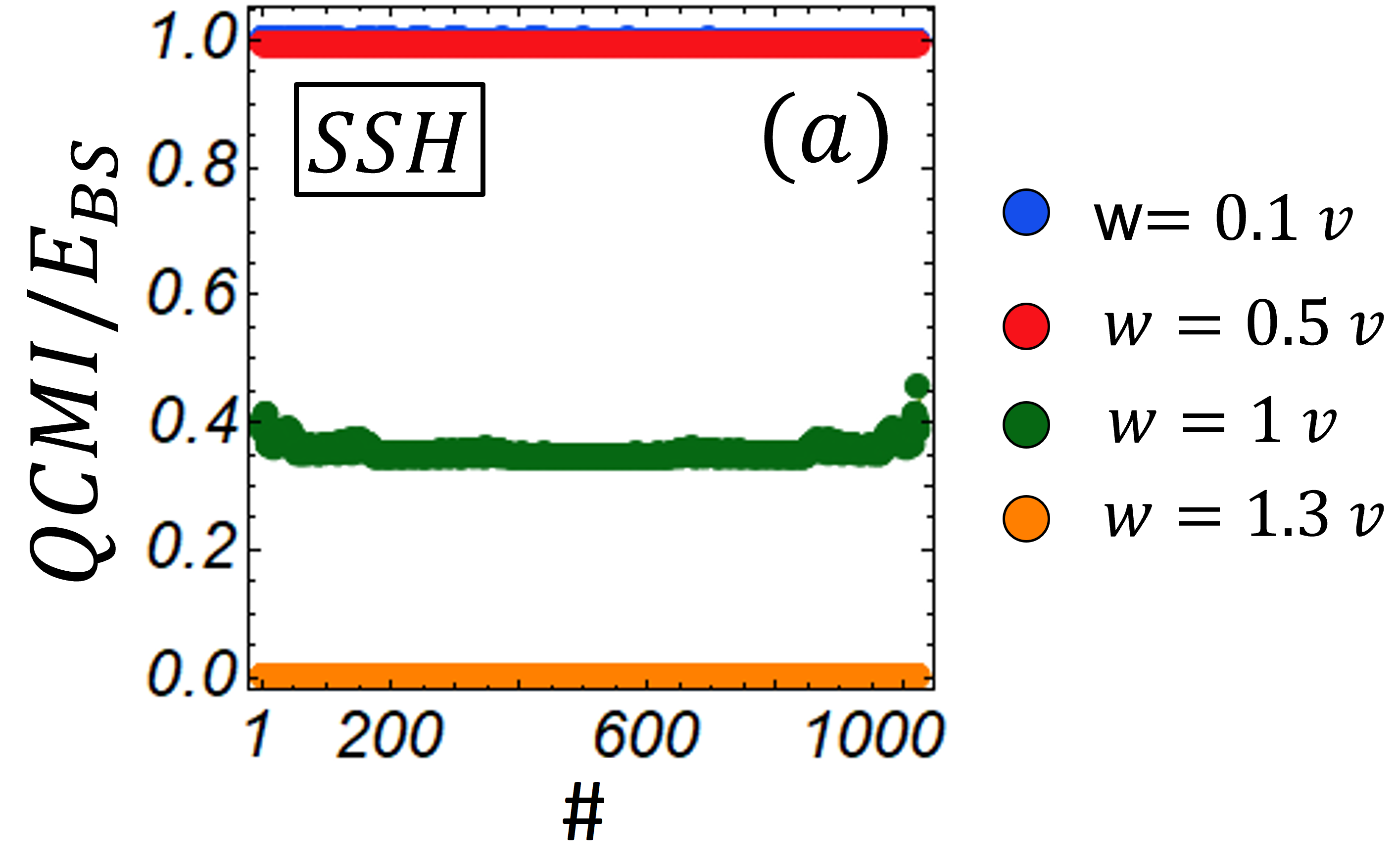}
	\hspace{0.5cm}
	\includegraphics[scale=0.12]{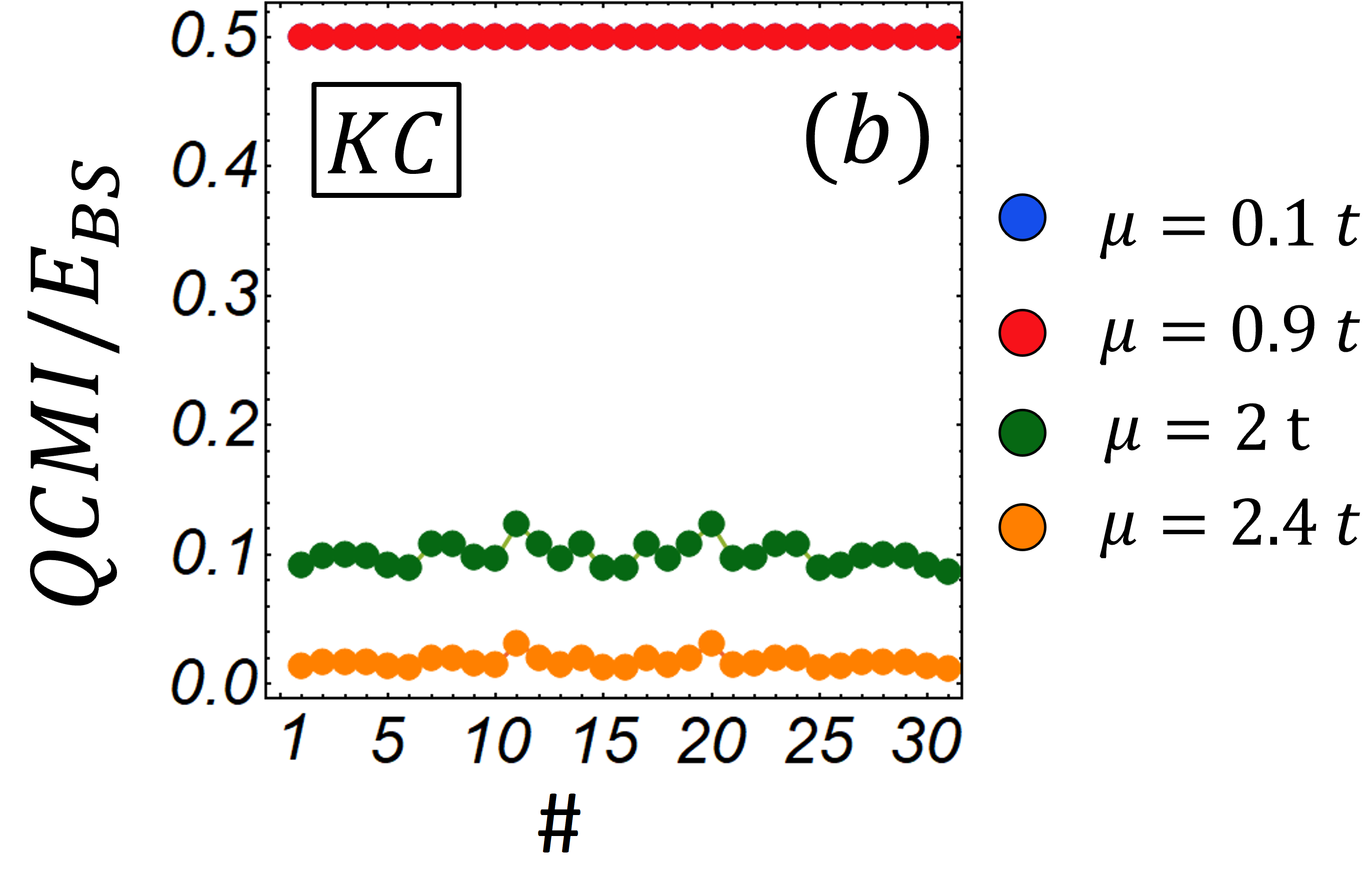}\\
	\vspace{0.5cm}
	\hspace{1.1cm}
	\includegraphics[scale=0.12]{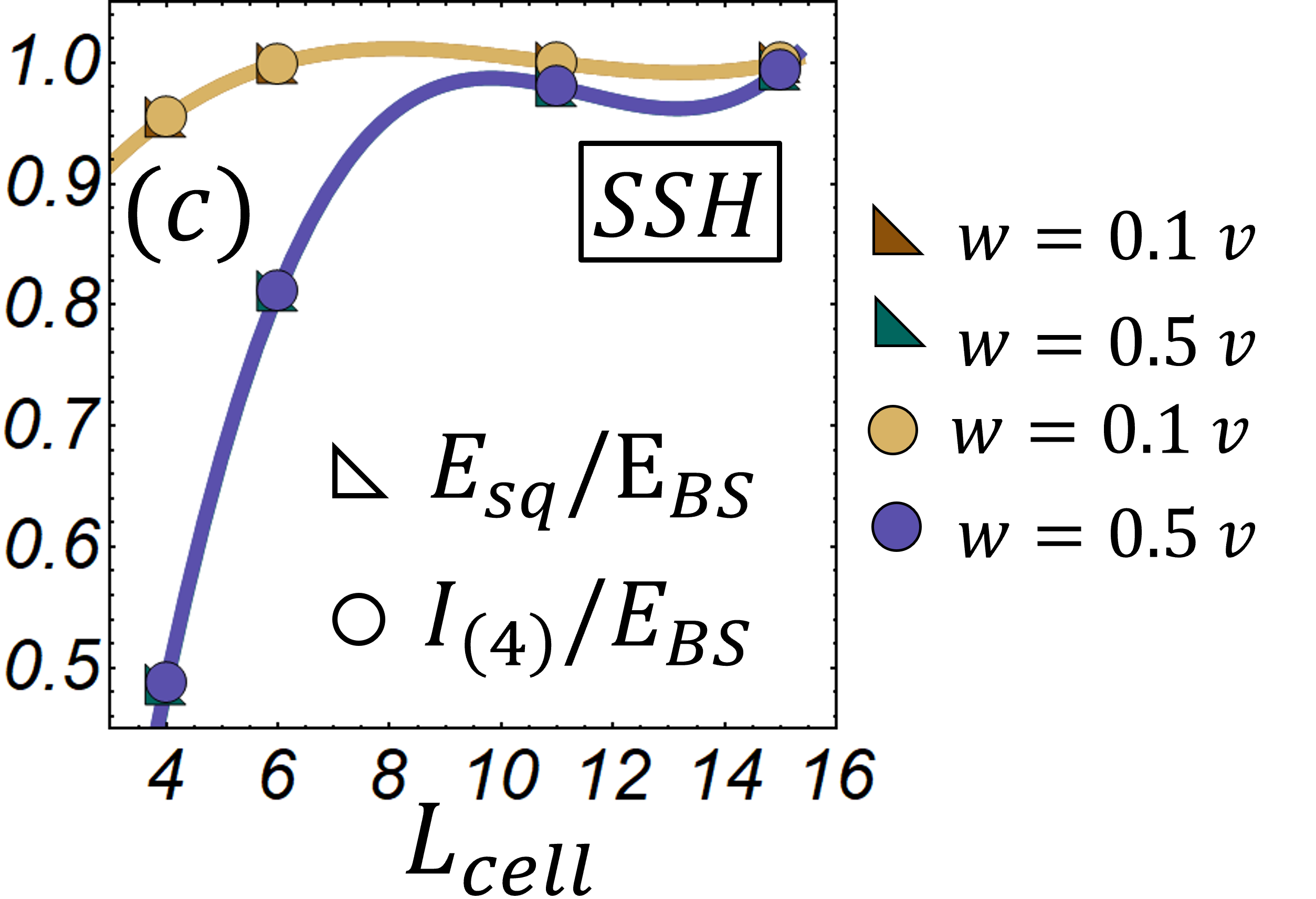}
	\hspace{1.3cm}
	\includegraphics[scale=0.12]{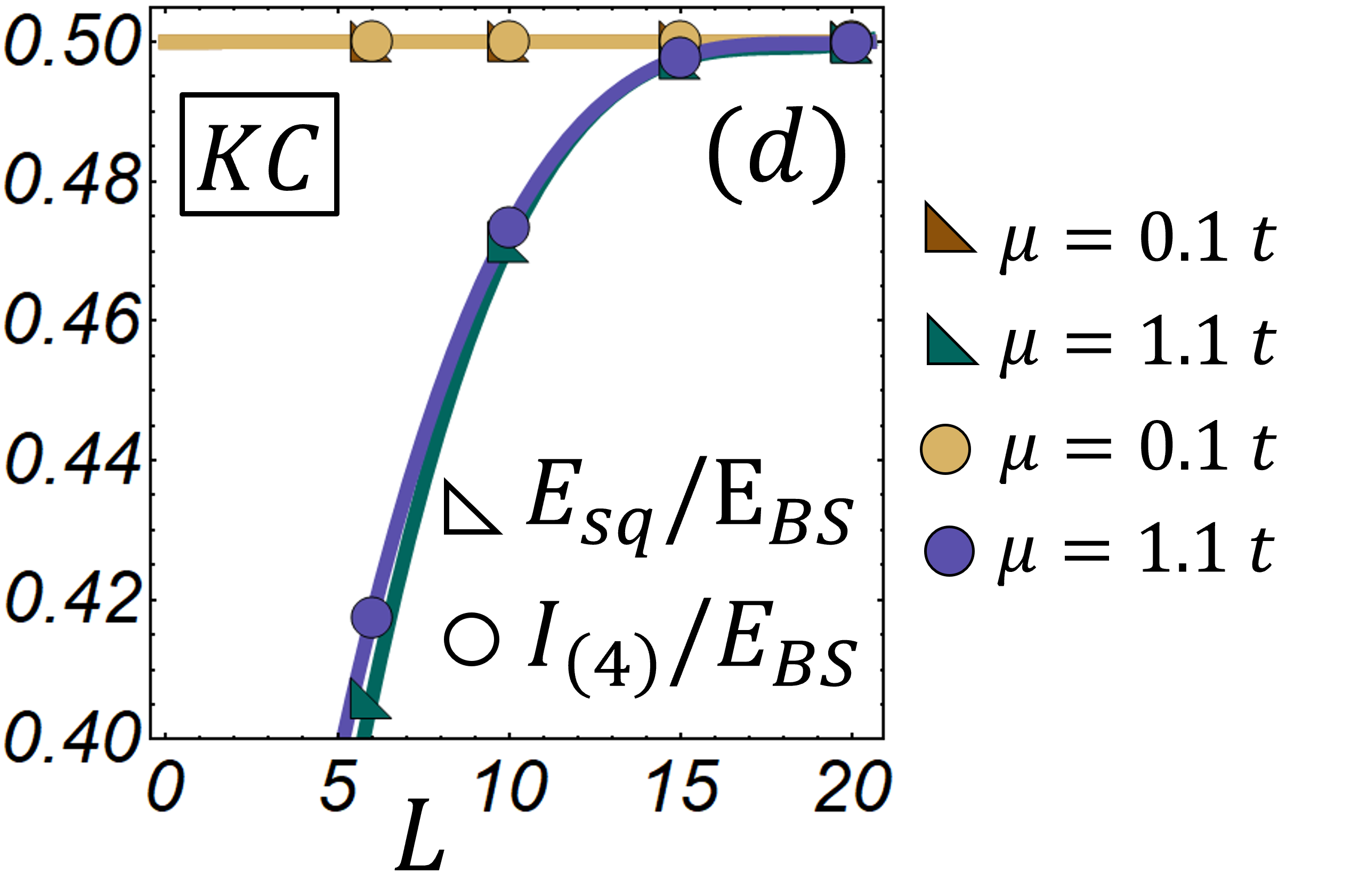}
	\hspace{0.2cm}
	\caption{(a)-(b): The edge-to-edge QCMI evaluated as a function of the number $\#$ of possible bulk quantum-state extensions $C$ inserted between edges $A$ and $B$ ($1\leq \# \leq N$). We have considered an SSH chain of $L_{cell}=15$ unit cells, panel (a), and a Kitaev chain of $L=15$ sites, panel (b). For the two systems, we fix the edge lengths respectively at $L_A=L_B\sim L_{cell}/3$ and at $L_A=L_B\sim L/3$. For the Kitaev chain we also set $t=\Delta$. The finite-size scalings of the approximate $E_{sq}$ and the exact $I_{(4)}$ are plotted in panels (c) and (d), respectively for the case of the SSH and of the Kitaev chain. In all cases, we find $E_{sq}=I_{(4)}$.}
	\label{appA}
\end{figure*}

Limited to a one-dimensional lattice of finite size $L$, the number $N$ of all the possible quantum-state extensions $Z$ is expressed by $N=\sum_{k=1}^{L_Z} \binom{L_Z}{k}$, $L_Z$ being the maximum size of $Z$ allowed, once the edge partitions have been fixed ($L_X$, $L_Y$), $L_Z=L-L_X-L_Y$. Since $N$ increases exponentially with $L_Z$ ($N=2^{L_Z}$), the computational resources required for the minimization of the QCMI explode exponentially as well, and the problem is indeed $NP$-hard. On the other hand, considering lattices of moderate size, one can determine an analytical approximant to the true SE restricted to one-dimensional quantum-state extensions. 

In Fig. \ref{appA}(a)-(b) we report the QCMI as a function of every one-dimensional quantum-state extension for four selected points of the phase diagram of the SSH chain, panel (a), and of the Kitaev chain, panel (b). One can see from Figs. \ref{appA}(a) and \ref{appA}(b) that for topological systems the actual value of the QCMI does not depend on the choice of the bulk partition. Indeed, topological states of matter represent special cases where the bulk is uncorrelated from the edges and the full, end-to-end QCMI is carried entirely by the edges alone. In this case, tracing out a connected or disconnected part of the bulk has no effects on the edge-to-edge correlations and all the possible different forms of the QCMI are expected to coincide. This is actually what occurs both in topological insulators and in topological superconductors. 

In Fig. \ref{appA}(c)-(d) we compare the finite-size scaling of the approximate SE $E_{sq}$ obtained via a succession of finite one-dimensional state extensions and the QCMI $I_{(4)}$ on a set of selected points of the topological phase diagram, respectively for the SSH model, see Fig. \ref{appA}(c), and for the Kitaev chain, see Fig. \ref{appA}(d). We observe that the QCMI $I_{(4)}$ coincides exactly with the approximate analytical expression obtained for the SE $E_{sq}$, but for a negligible numerical discrepancy originated by finite-size effects. For a topological insulator of moderate system sizes the approximate analytical expression of the SE $E_{sq}$ between the system edges converges to the Bell state entanglement: $E_{sq}(\rho_{AB})=\ln 2$, while for a topological superconductor it converges to half the Bell state entanglement: $E_{sq}(\rho_{AB})=\ln 2/2$.

Hereafter, we provide an approximate analytical argument that $E_{BS}/2$ represents the bulk limit of the edge-to-edge SE in the Kitaev superconducting chain. First of all, the constant quantized value $E_{BS}/2$ taken by the QCMIs can be analytically determined at the points of exact ground-state topological degeneracy, i.e. at $\mu=0$, $t=\Delta$. In this limit, the topological edge modes decouple from the bulk and all QCMIs become independent of the chain size. Due to the above properties, $I_{(3)}$ and $I_{(4)}$ are expected to coincide with the true SE without size constraints, see Fig. \ref{appA}. For this reason, without loss of generality, we consider the collection of the many-body ground-state density matrices $\rho_{ABC}$ of the tripartite system $ABC$ as the set of possible state-extensions for the approximate computation of the edge-to-edge SE.

A Kitaev chain Hamiltonian with $\mu=0$ and $t=\Delta$ can be expressed via the Jordan-Wigner mapping as $H_{spin}^{KC}= -t\sum_{j=1}^{L-1} \sigma^x_{j} \sigma^x_{j+1}$. The density matrix of the parity-preserving many-body ground state reads:
\begin{eqnarray}
	\begin{split}
		\rho_{ABC}=&\frac{1}{2}\ \biggl[ \ket{\uparrow \dots \uparrow}\!\!\bra{\uparrow \dots \uparrow}_L+\ket{\uparrow \dots \uparrow}\!\!\bra{\downarrow \dots \downarrow}_L+\\ 
		&\ket{\downarrow \dots \downarrow}\!\!\bra{\uparrow \dots \uparrow}_L+\ket{\downarrow \dots \downarrow}\!\!\bra{\downarrow \dots \downarrow}_L\biggr] \, ,
	\end{split}
\end{eqnarray}
where $L$ is the length of the chain and the notation $\ket{\alpha \dots \alpha}\!\!\bra{\beta \dots \beta}_L$ has been introduced to denote a matrix with $L$ spins where $\ket{\alpha} / \ket{\beta}=\ket{\uparrow}, \ket{\downarrow}$ and with $\ket{\uparrow}=1/\sqrt{2}\ (1,1)^T$ and $\ket{\downarrow}=1/\sqrt{2}\ (1,-1)^T$. It is straightforward to show that, independently on the partition lengths, tracing out a part of the system leads to the following reduced density matrices:
\begin{align}
	\nonumber
	&\rho_{AC}\!=\!\frac{1}{2}\biggl[ \ket{\uparrow \dots \uparrow}\!\!\bra{\uparrow \dots \uparrow}_{L_A+L_C}\!\!+\!\ket{\downarrow \dots \downarrow}\!\!\bra{\downarrow \dots \downarrow}_{L_A+L_C}\!\!\biggr],\\ \nonumber
	\vspace{0.5cm}
	&\rho_{BC}\!=\!\frac{1}{2}\biggl[ \ket{\uparrow \dots \uparrow}\!\!\bra{\uparrow \dots \uparrow}_{L_B+L_C}\!\!+\!\ket{\downarrow \dots \downarrow}\!\!\bra{\downarrow \dots \downarrow}_{L_B+L_C}\!\! \biggr],\\
	\vspace{0.5cm}
	&\rho_{C}\!=\!\frac{1}{2}\biggl[ \ket{\uparrow \dots \uparrow}\!\!\bra{\uparrow \dots \uparrow}_{L_C}\!\!+\!\ket{\downarrow \dots \downarrow}\!\!\bra{\downarrow \dots \downarrow}_{L_C} \biggr] \, ,
	\label{Exvon}
\end{align}
where $L_A$, $L_B$ and $L_C$ refer respectively to the lengths of the edges $A$, $B$ and of the bulk $C$. Via the expressions in Eq. \ref{Exvon}, the reduced von Neumann entropies are easily computed. It turns out that $S_{AC}=S_{BC}=S_{C}=\log2$, while of course for the pure ground state projector $S_{ABC}=0$. Therefore, the end-to-end (edge-edge) SE of the reduced two-edge state $\rho_{AB}$ reads $E_{sq}(\rho_{AB})=\log 2/2$. A similar derivation holds for the SSH chain with $w=0$, $v \neq 0$, leading to $E_{sq}(\rho_{AB})=\log 2$.

In conclusion, we have shown how to compute an approximate form of the SE between the edges restricted only to one-dimensional state extensions and for systems of finite size, and we have found that it coincides with the two distinct forms $I_{(3)}$ and $I_{(4)}$ of the QCMI between the edges. It turns out that he approximate SE saturates to the bulk value expected for the true edge-to-edge SE already for systems of limited size. We have shown that the aforementioned bulk value can be explicitly computed according to the analytical proof provided above. In view of these observations, we conclude that the conjecture $E_{sq}=I_{(3)}=I_{(4)}$ is strongly corroborated and should hold without restrictions on the system size because the bulk part of the system does not contribute to the quantum correlations between the edges within topologically ordered phases. %This latter observation suggests that the numerical values of $I_{(3)}$ and $I_{(4)}$ are indistinguishable %from the exact value of the genuine ground-state SE for any system size. Accordingly, in the main text we %identify the numerical values of $I_{(3)}$ or $I_{(4)}$ with the true edge-to-edge SE.

\section{Discussion}
\label{SecV}
In the present work we have shown that the squashed entanglement between the edges of a quantum many-body system, together with its natural upper bounds defined by two inequivalent forms of the edge-to-edge quantum mutual information conditioned by the system bulk, realize the natural framework of non-local order parameters characterizing topological quantum matter in one dimension. Such quantities discriminate between topological insulators and superconductors by identifying the different physical nature and statistics of the edge modes in the two cases. Use of the edge-to-edge quantum conditional mutual information yields the exact quantum phase diagram for the non-interacting models; in the presence of interactions, it reproduces both qualitatively and quantitatively the consensus results obtained so far by various analytical and numerical approaches. The latter highly nontrivial result makes the quantum conditional mutual information between the system edges the natural benchmark to test the accuracy of numerical approximations. 
%We have compared the squashed entanglement, computed by means of $I_{(3)}$ and $I_{(4)}$, with the ill-%defined but albeit very popular entanglement negativity, showing how the latter fails in reproducing the %correct phase diagram of topological superconductors, and thus yielding further support to the paradigm of %the multipartition-based measure of subsystem bipartite entanglement.

The fundamental property of the quantum conditional mutual information and squashed entanglement that singles them out when compared to standard quantifiers of pure-state bipartite entanglement such as the block entanglement entropy and the entanglement spectrum is that the former, although being bipartite measures, are defined, at variance with the latter, in terms of multipartitions. This property in turn warrants that quantum conditional mutual information and squashed entanglement are sensitive to and can detect the different physical nature of the bulks (e.g. conducting or insulating) and of the edges for different systems and can thus distinguish and discriminate topologically ordered phases from ordinary Ginzburg-Landau symmetry-breaking orders as well as identify and discern different topologically ordered regimes. In essence, this fine-grain ability keeps track of the different parts of a system, either edge-bulk-edge or edge-partial bulk-partial bulk-edge, and it is responsible for the tremendous effectiveness of edge-state quantum conditional mutual information and edge-state squashed entanglement in detecting, identifying and discriminating different types of quantum phase transitions and collective quantum orders.

A crucial step in order to extend our results to include all topological quantum matter is to generalize the concept of edge-state quantum conditional mutual information and edge-state squashed entanglement to generic many-body systems in dimension $D \geq 2$. For such higher-dimensional cases, the main challenge lies in the correct identification of the appropriate bulk and edge parts for a given multipartition. Among two-dimensional systems, second order topological materials ($HOTM_2$) \cite{PhysRevLett.122.236401,PhysRevResearch.2.033495} play a fundamental role, both as novel platforms for topological insulators or to realize topological superconducting braiding dynamics \cite{PhysRevResearch.2.032068}. $HOTM_2$ are systems with gapped one-dimensional boundaries and zero-dimensional localized modes (corner states). For such systems, the identifications of edges $A$, $B$ and bulks
$C$ $C_1$, $C_2$ is clear and easily connected to that of the one-dimensional case. This class of systems is thus the first prominent candidate in order to implement and test the scheme of topological squashed entanglement to many-body systems in $D=2$.

Concerning the experimental accessibility of topological squashed entanglement, the problem boils down to that of measuring quantum entropies of a set of reduced states. A recent proposal relies on the thermodynamic study of the entanglement Hamiltonian for the direct experimental probing of von Neumann entropies via quantum quenches \cite{Mendes_Santos_2020}. A possibility specifically taylored for systems featuring topological order consists in identifying minimum entropy states and then experimentally simulating the behaviour of the associated von Neumann entropies via the classical microwave analogues of such states \cite{Zhang_2019}. A further intriguing possibility arises from the observation that highly informative bounds on von Neumann entropies, quantum conditional mutual information, and squashed entanglement can be constructed in terms of R\'enyi entropies~\cite{Brandao2015,FawziRenner2015}. The strategy is then to adapt to fermionic systems~\cite{Cornfeld2019} the schemes previously proposed for the experimental probe of R\'enyi entropies in bosonic and spin systems~\cite{Abanin2012,Daley2012,Elben2018} and the corresponding techniques that led to the first measurement of the quadratic R\'enyi entropy in a many-body system~\cite{Islam2015}.

\appendix
\section{Finite-size scaling}
 \label{AppA}
As already showed SE scales exponentially to the entanglement shared by the non-trivial edge modes in the topological ordered phases, while it is
pinned to zero for points out of the such regime. At phase boundaries the edge modes are less localized at the ends of the chain and possible nontrivial effects on SE should appear more transparent.
\begin{figure}
	\includegraphics[scale=0.15]{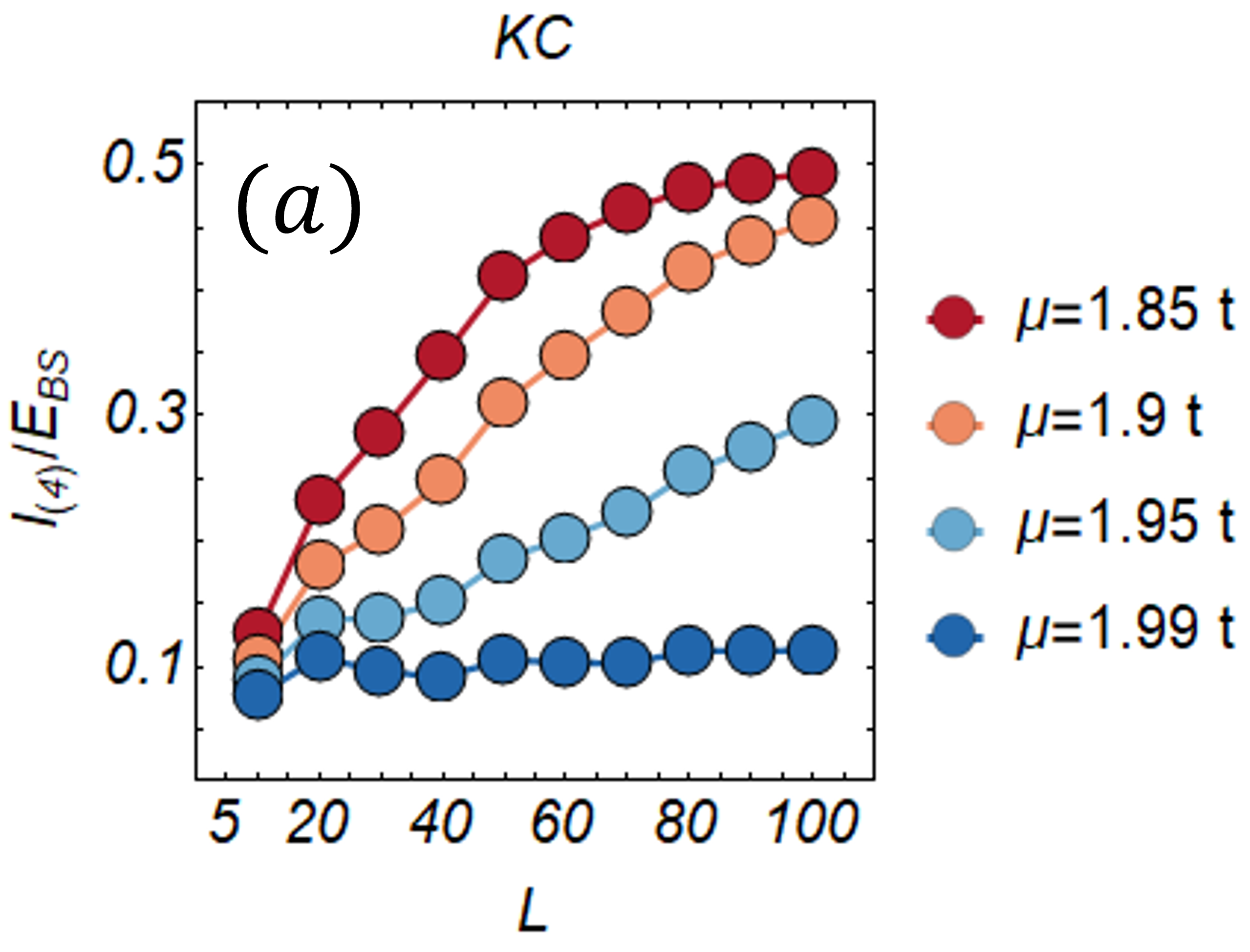}
	\includegraphics[scale=0.15]{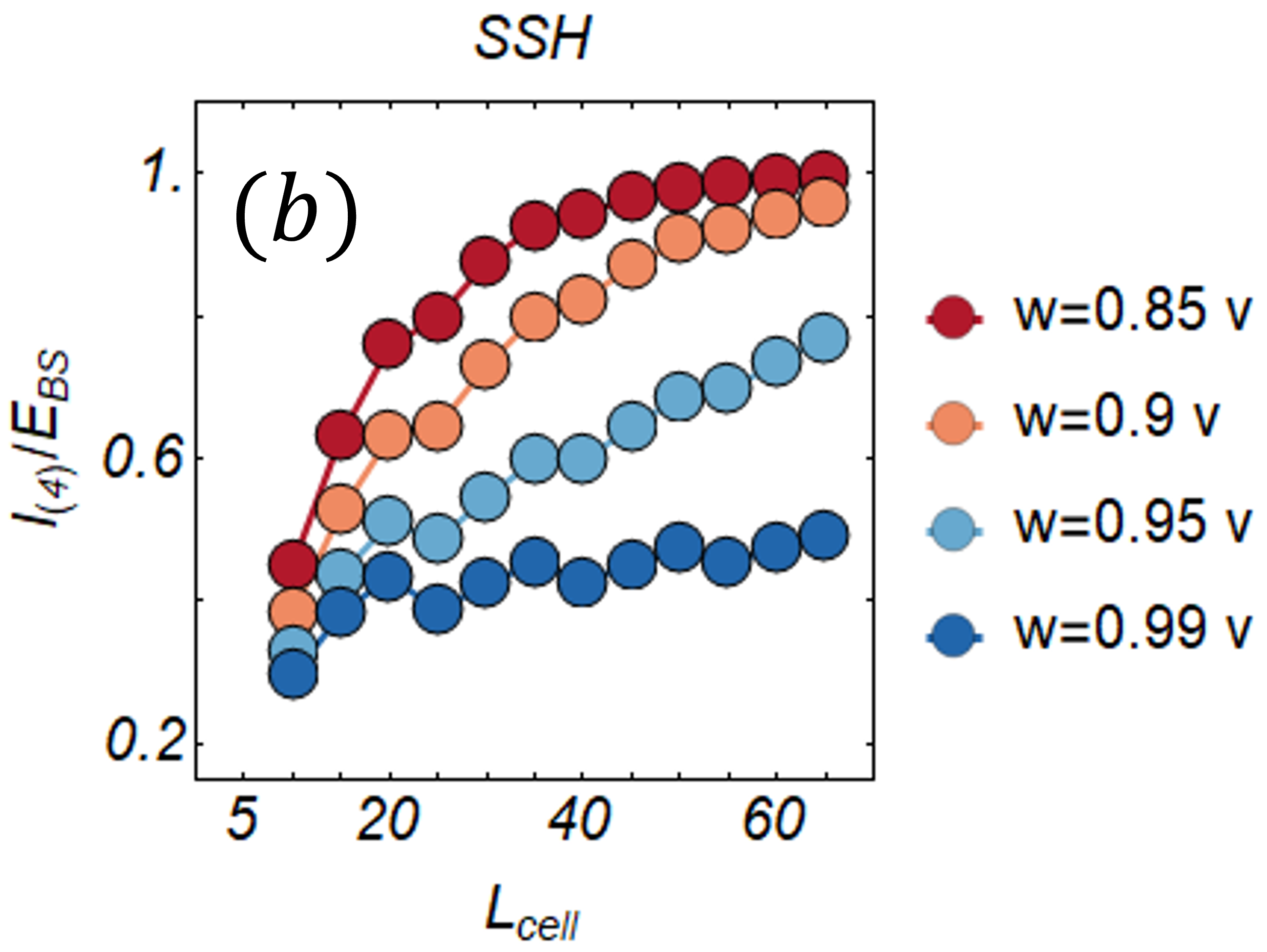}
	\caption{Finite-size scaling at phase transition boundaries of the QCMI to Bell-state entanglement ratio $I_{(4)}/E_{BS}$ for the SSH chain  (a) and the KC chain (b).}
	\label{Figure6}
\end{figure}
In panels (a) and (b) of Fig. \ref{Figure6} we show that, for fixed size, the QCMI to Bell-state entanglement ratio $I_{(4)}/E_{BS}$ is lowered when topological phase transition points are approached: $w=v$, $\mu=2t$. However an increasing scaling with size towards $E_{BS}$ and $E_{BS}/2$ respectively for the SSH chain, panels (a), and the Kitaev model, panel (b), is observed, signaling the robustness of the entanglement non-local order parameter even close to the phase transition. Significant deviations from the quantized entanglement values are only obtained when $w$ and $t$ differ form aforementioned phase transition points by the order of $10^{-2}$.

\section{Effects of disorder}
\label{AppB}
Robustness against disorder is a specific property of topological materials \cite{Brouwer2011,PhysRevMaterials.5.014204,PhysRevB.104.195117,Hegde2016}. In fact, disorder may even induce localization effects as in the case of Anderson insulators  \cite{PhysRevLett.102.136806}, thus favouring topological phases of matter, whereas in other systems such as, e.g., semiconductor-based Majorana nanowires and topological insulator nanoribbons, it can yield detrimental effects \cite{PhysRevMaterials.5.014204,PhysRevMaterials.5.124602}. The topologically ordered phases of the SSH and Kitaev chains are robust to the effects of disorder and local perturbations \cite{Brouwer2011,Neven2013,Adagideli2014,Hegde2016,10.21468/SciPostPhysCore.3.2.012}.

\begin{figure}
	\includegraphics[scale=0.075]{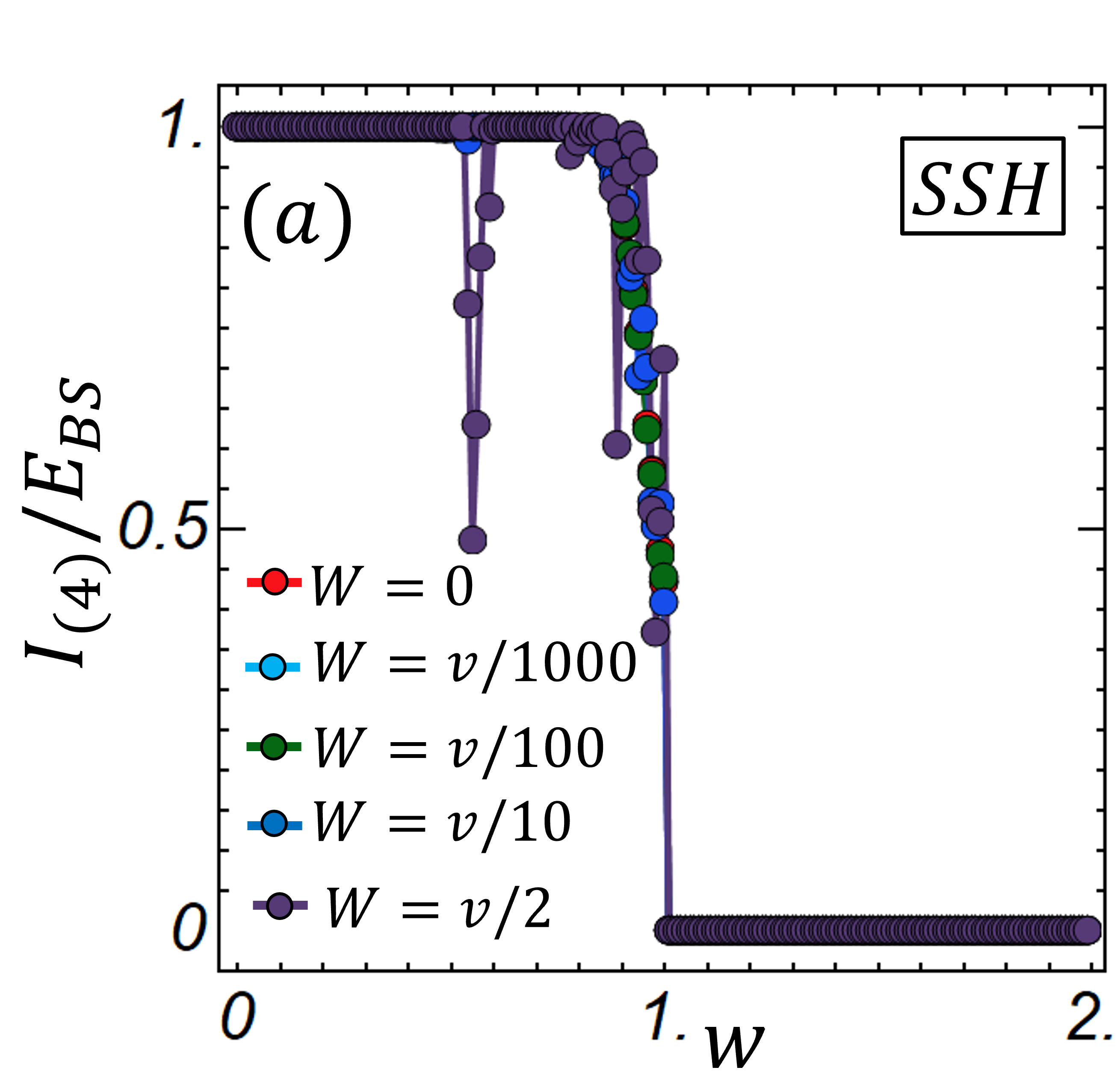}
	\includegraphics[scale=0.075]{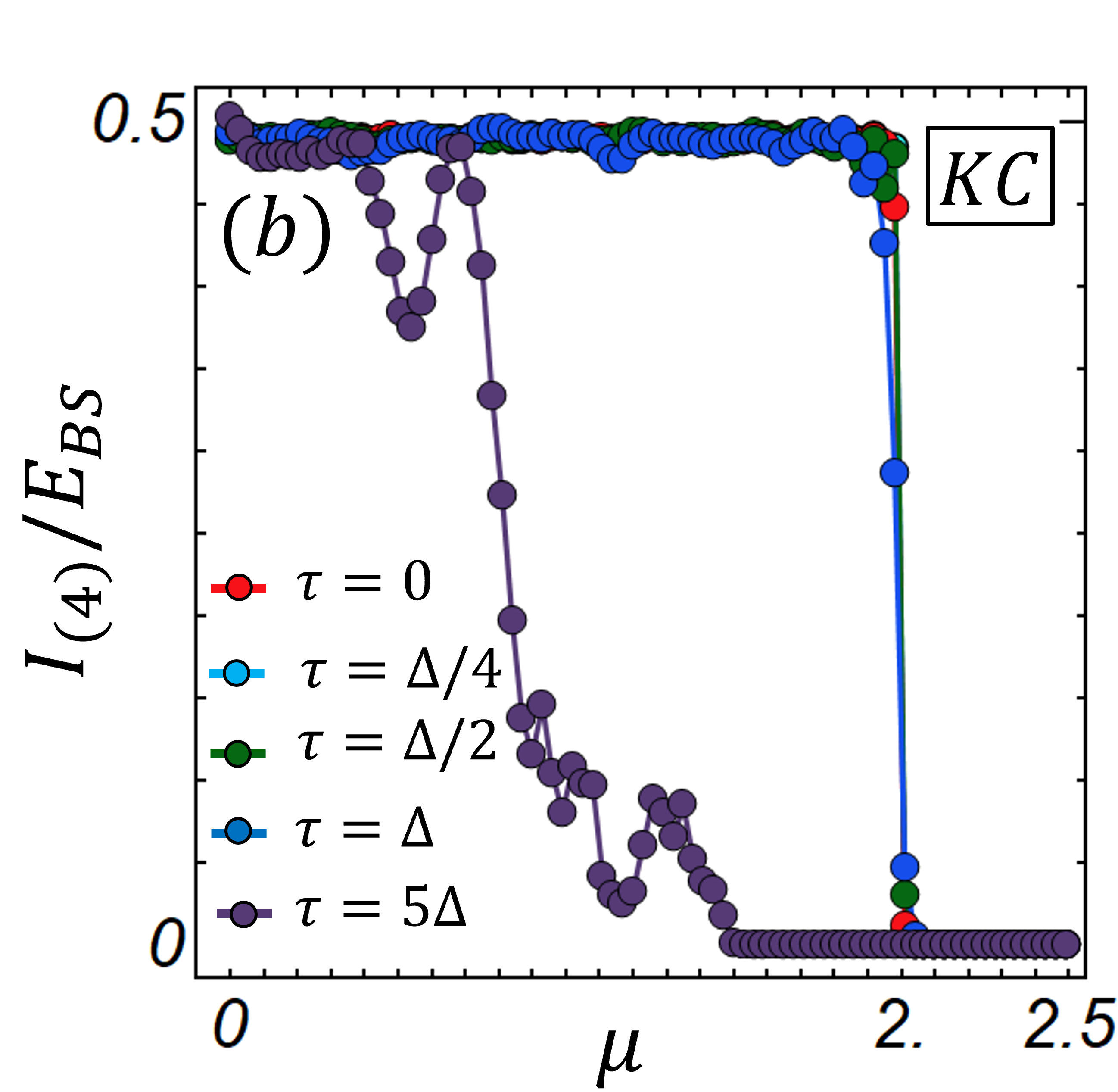}
	\caption{The QCMI to Bell-state entanglement ratio $I_{(4)}/E_{BS}$ in the presence of disordered hopping integrals in the SSH and Kitaev chains. Panel (a): behaviour of $I_{(4)}/E_{BS}$ for the SSH chain with random hopping amplitudes $w_i=w+W_1 \delta_i$ and $v_i=v+W_2 \delta_i$, with $\delta_i$ randomly generated in the interval $(-0.5, 0.5)$ and with $2W_1=W_2=W$, $v=1$, $L_{cell}=50$. Panel (b): behaviour of $I_{(4)}/E_{BS}$ for The Kitaev chain with random hopping amplitudes $t_i$. The random realizations are generated by a uniform probability distribution defined in the interval $t_i \in (t-\tau, t+\tau)$. The reference values of the parameters have been fixed as $t=1$, $\Delta=0.1$, $L=50$.}
	\label{Figure7}
\end{figure}

For both models, we study the response of the QCMI to the disorder induced on the hopping integrals. This choice provides a model of the effective mass gradient and random doping along the system that originate from the growth process of the one-dimensional nanowires. Specifically, we consider uniform, chirality-preserving disorder on the hoppings of the SSH chain:
\begin{eqnarray}
	w_i=w+W_1 \delta_i \, , \\
	v_i=v+W_2 \delta_i \, ,
	\label{DisSSH}
\end{eqnarray}
with $2W_1=W_2=W$, and $\delta_i$ randomly generated in the interval $(-0.5,0.5)$. For the Kitaev chain we consider the following random hopping integrals:
\begin{eqnarray}
	t_i=t+\tau_i^{dis} \, ,
\end{eqnarray}
with $\tau_i^{dis}$ uniformly distributed in the interval $(-\tau,\tau)$.

In Fig. \ref{Figure7}, we let $W$ and $\tau$ range from perturbative values $W=v/10000$ and $\tau=\Delta/4$ up to a regime for which the strength of the disorder becomes comparable with the band gaps of the two models, respectively $W=v/2$ and $\tau=5\Delta$. 

For the SSH chain, we see that the QCMI to Bell-state entanglement ratio $I_{(4)}/E_{BS}$ remains resilient over all the examined regimes, only being affected by some fluctuations in the regime of very high disorder, especially near the phase boundary. A similar phenomenology holds for the superconducting chain from $\tau=\Delta/4$ up to $\tau=\Delta$. For higher values of $\tau$, e.g. when $\tau=5 \Delta$ the phase boundary defining the topological transition tends to be suppressed. This effect suggests that random hopping is more effective in perturbing the topologically ordered phase in the Kitaev chain then in the SSH chain. At any rate, the above analysis yields that realistic values of the disorder strength do not cause any significant disruption of the topologically ordered phases as measured by the SE.

\section{Interacting systems}
\label{AppC}
\begin{figure}
	\includegraphics[scale=0.08]{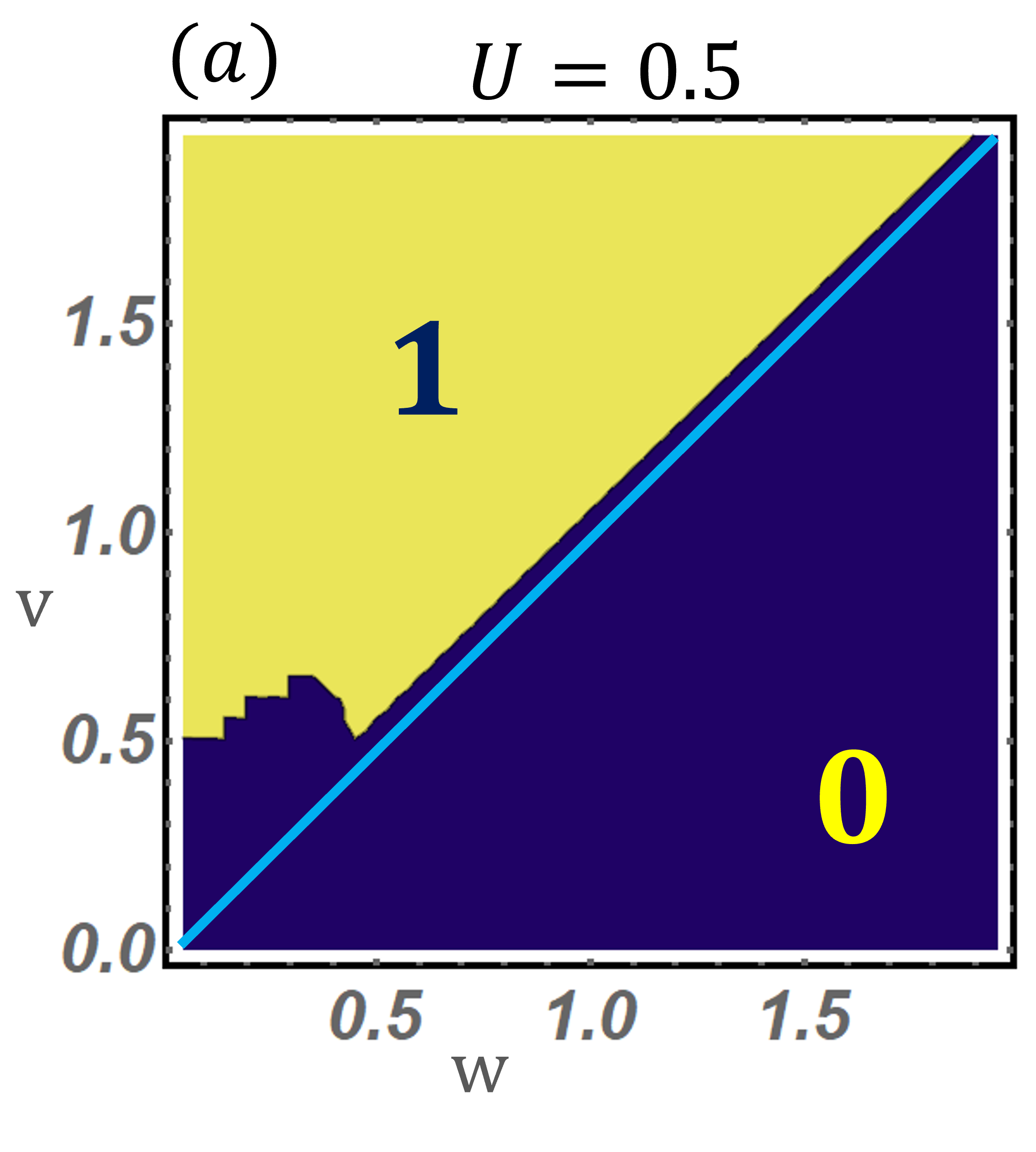}
	\includegraphics[scale=0.08]{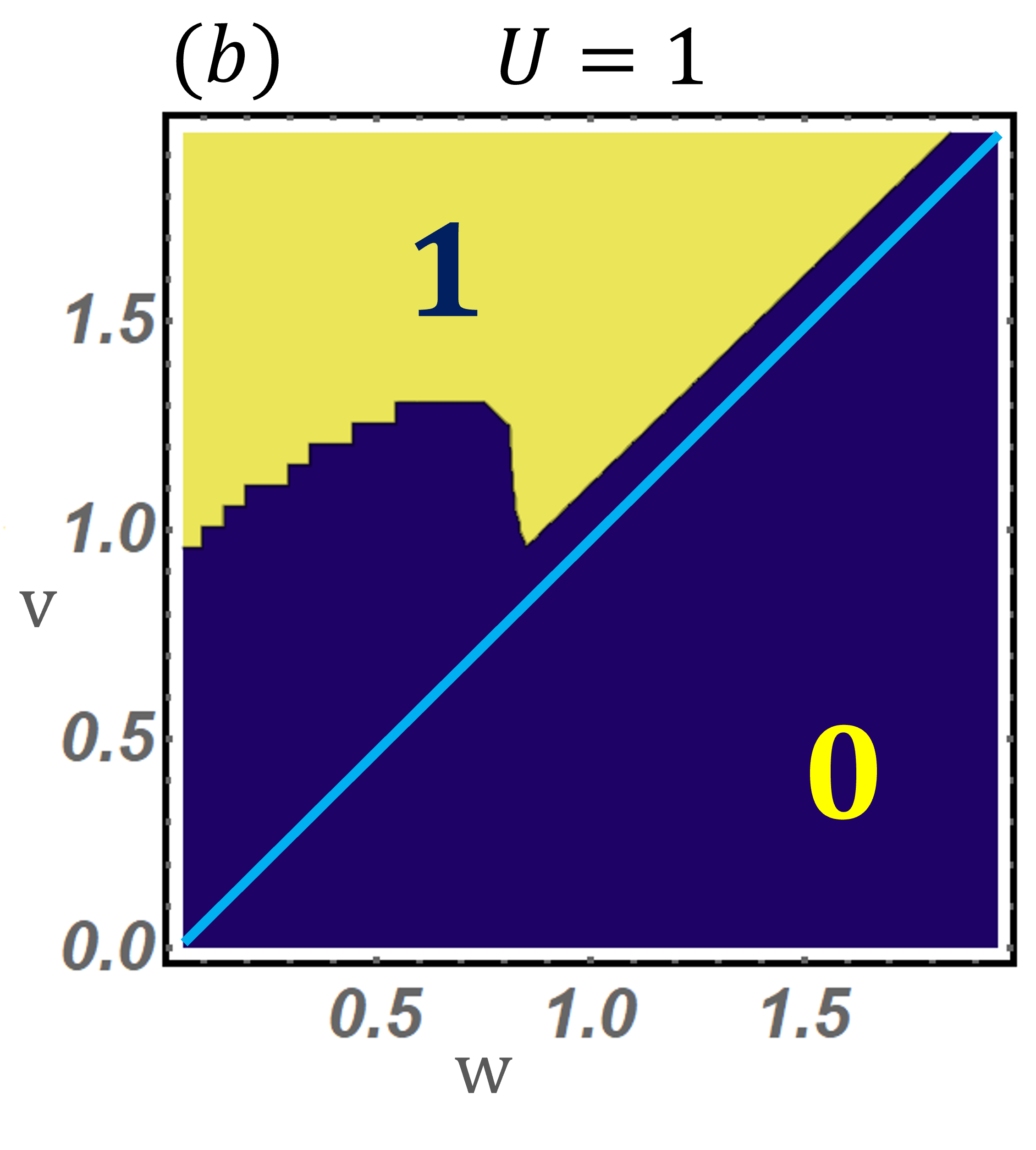}
	\includegraphics[scale=0.08]{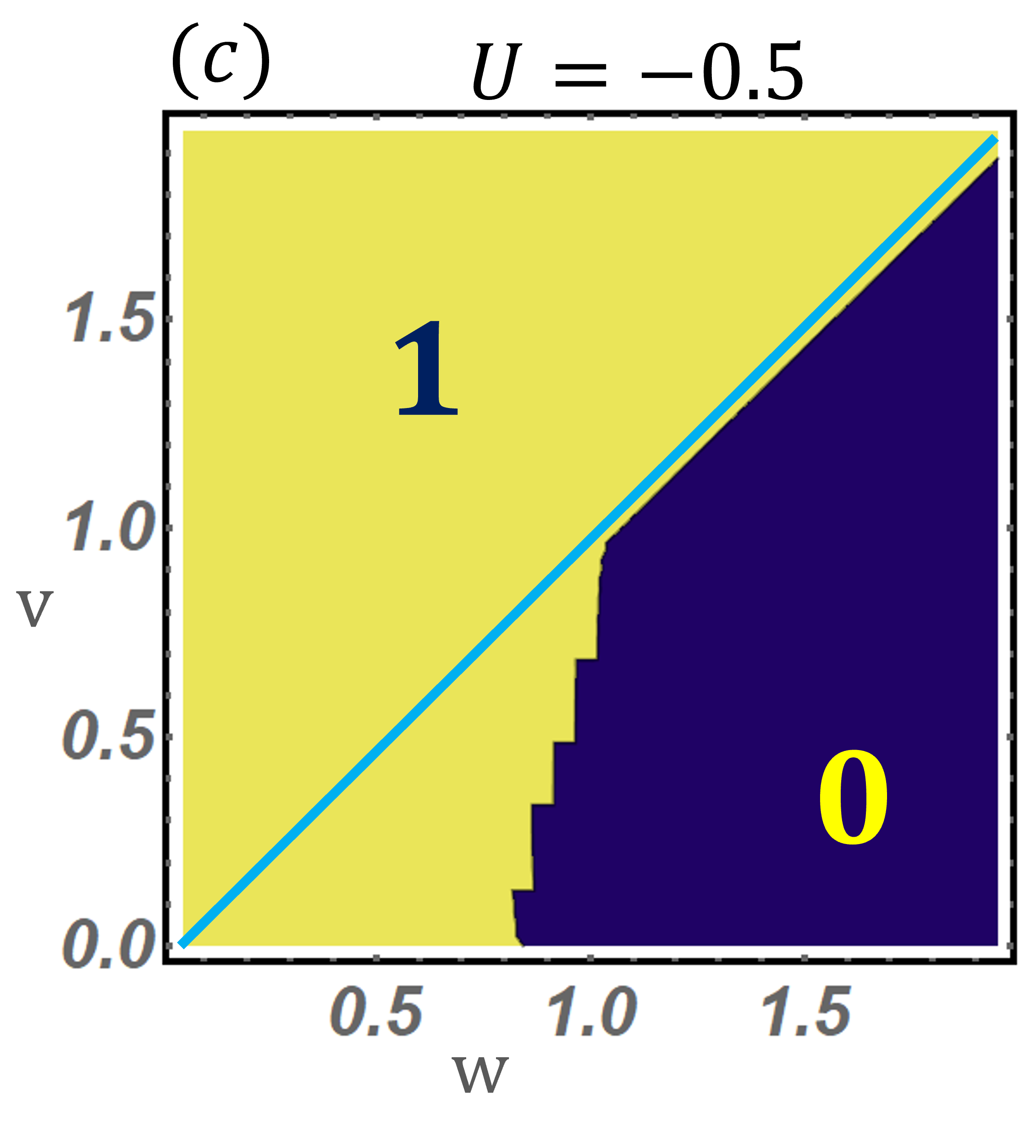}
	\includegraphics[scale=0.08]{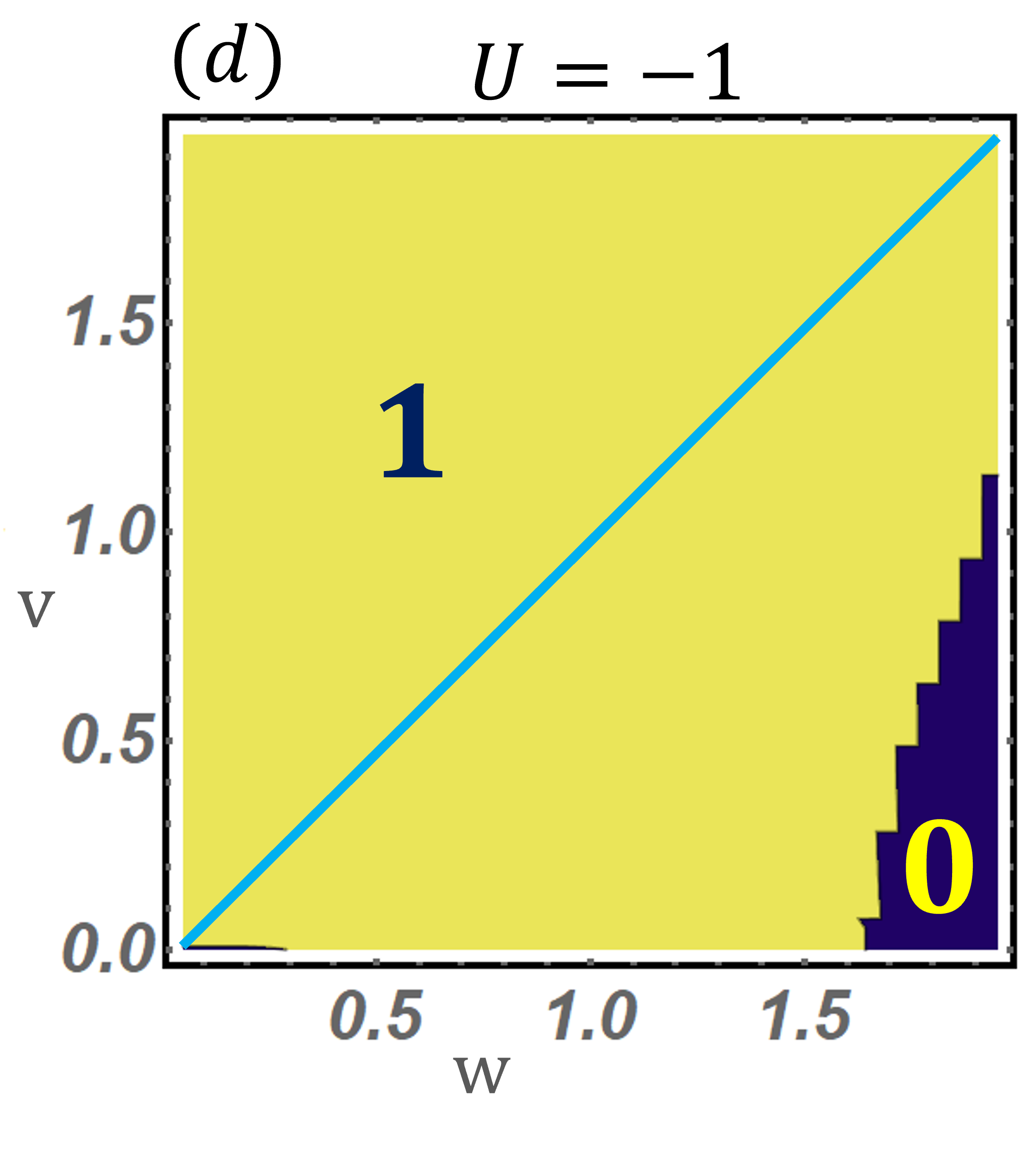}	
	\caption{Phase diagrams of the interacting SSH chain in the $w-v$ plane at different fixed values of the interaction strength $U$. The light blue lines correspond to the phase boundaries of the non-interacting case ($w=v$). The contour plots reproduce the behavior of the function $Sign[(I_{(3)}-I_{(3)}^{w=v})/E_{BS}$], where $I_{(3)}^{w=v}/E_{BS}=0.15$ is the value of $I_{(3)}$ when $U=0$ and $w=v$. The length of the chain is fixed at $L_{cell}=14$.}
	\label{Figure8}
\end{figure}
\subsection{Mapping to interacting spin Hamiltonians}
The Jordan-Wigner transformation \cite{Franchini2017} is an highly non-local mapping between fermionic operators and spin $1/2$ operators. On each site, an empty state is mapped into a spin up and an occupied one to a spin down. The non-local part of this mapping is called the Jordan-Wigner string and fixes the (anti)commutation relation between sites, by counting the parity of overturned sites to the left of the spin on which it is applied. 

This transformation explicitly breaks the translational invariance of the model, by singling out a particular site as a starting point for the string. Denoting by $c_{j,\alpha}$ and $c_{j,\alpha}^\dagger$ the generic annihilation and creation fermionic operators, the Jordan-Wigner mapping is defined by:

\begin{eqnarray}
	%\begin{cases}
	%\vspace{0.1cm}
	c_{j,\alpha} &=& e^{-i \pi \sum_{\alpha'} \sum_{l=1}^{j-1} c^\dagger_{l,\alpha'} c_{l,\alpha'}}\sigma^{+}_{j,\alpha} \, ,\\
	&& \nonumber \\
	c^\dagger_{j,\alpha} &=& \sigma^{-}_{j,\alpha}e^{i \pi \sum_{\alpha'} \sum_{l=1}^{j-1} c^\dagger_{l,\alpha'} c_{l,\alpha'}} \, , \\
	&& \nonumber \\
	n_{j,\alpha} &=& \frac{1-\sigma^{z}_{j,\alpha}}{2} \, ,
	%\end{cases}
	\label{JWtransformation}
\end{eqnarray}
where $j$ singles out the explicit lattice site while $\alpha$ denotes the remaining degrees of freedom of the system. The aforementioned parity string of the overturned sites is $e^{-i \pi \sum_{\alpha'} \sum_{l=1}^{j-1} c^\dagger_{l,\alpha'} c_{l,\alpha'}}$. The operators $\sigma_{j,\alpha}^{(+,-)}=(\sigma_{j,\alpha}^{x}\pm i\sigma_{j,\alpha}^{y})/2$ are the well-known combination of Pauli matrices and the last relation in Equation (\ref{JWtransformation}) allows to express the parity operator of the fermionic site $j$ with degrees of freedom $\alpha$ as $e^{-i \pi c^\dagger_{j,\alpha} c_{j,\alpha}}=\sigma_{j,\alpha}^{z}$. 

Using the algebra of spin $1/2$ operators and the constraint that on different sites Pauli matrices commute, it is straightforward to derive the following spin-$1/2$ representations of the interacting SSH insulator and Kitaev superconductor:
\begin{eqnarray}
	H_{spin}^{SSH}=&&\frac{1}{2} \biggr [U \bigl(L-\frac{1}{2}\bigr) + w \sum_{j=1}^{L} \bigl(\sigma^x_{aj} \sigma^x_{bj}+\sigma^y_{aj} \sigma^y_{bj}\bigr) + \nonumber \\
	&&+v\sum_{j=1}^{L-1}\bigl(\sigma^x_{bj} \sigma^x_{aj+1}+\sigma^y_{bj} \sigma^y_{aj+1}\bigr)+\frac{U}{2}\bigl(\sum_{1}^{L} \sigma^z_{aj} \sigma^z_{bj} + \nonumber \\
	&&+\sum_{1}^{L-1}\sigma^z_{bj} \sigma^z_{aj+1}\bigr)-U \bigl( \sum_{j=2}^{L} \sigma^z_{aj}+\sum_{j=1}^{L-1}\sigma^z_{bj}\bigr) + \nonumber \\
	&&-\frac{U}{2}\bigl(\sigma^z_{a1}+\sigma^z_{bL}\bigr) \biggl] \, ,
	\label{JWSSH}
\end{eqnarray}

\begin{eqnarray}
	\begin{split}
	H_{spin}^{KC} \!=& \!\!\sum_{j=1}^{L-1} \!\biggl(-(t\!+\!\Delta) \sigma^x_{j} \sigma^x_{j+1}\! +\! (t \!- \!\Delta)\sigma^y_{j} \sigma^y_{j+1} + U \sigma_j^z\sigma_{j+1}^z\!\biggr) \\
&+\frac{\mu}{2} \sum_{j=1}^{L} \sigma^z_{j} \, .
	\label{JWKC}
\end{split}
\end{eqnarray}

When the SSH Hamiltonian in Eq. \ref{JWSSH} is considered, $\alpha=a$, $b$ represent the degree of freedom of the unit cell, while the Kitaev chain Hamiltonian in Eq. \ref{JWKC} describes exactly one fermionic degree of freedom per lattice site. 
We see that the interacting SSH model transforms into a staggered $XYZ$ chain with external magnetic field along the $z$-direction, while the Kitaev chain is described by a $XYZ$ chain with external magnetic field along the $z$-direction. Moving to the spin representation, the Hilbert space dimension grows exponentially from $2L$ to $2^L$.

\subsection{Phase diagram of the interacting SSH chain}
By means of the Jordan-Wigner mapping, the edge-to-edge QCMI can be computed exactly for interacting systems of relatively modest size. In the main text we have discussed the phase diagrams of the interacting SSH and Kitaev models by means of the $I_{(3)}$ in the $U-w$ and $U-\mu$ plane of the phase space. Here we discuss the phase diagram of the interacting SSH chain in the $w-v$ plane. 

In Fig. \ref{Figure8} (a)-(d) we report four different phase contour plots, each corresponding to a fixed value of the Coulomb interaction strength $U$. 
As the Coulomb repulsion $U$ is increased, the topological phase boundary gets progressively reduced in comparison to that of a non interacting case $w=v$ (blue light line). On the other hand, for increasing Coulomb attraction the topological phase reopens, becoming the dominant contribution of the entire phase diagram.

\section*{Acknowledgements}
F.I. and A.M. acknowledge support by MUR (Ministero dell’Università e della Ricerca) via the project PRIN 2017 ”Taming complexity via QUantum Strategies: a Hybrid Integrated Photonic approach” (QUSHIP) Id. 2017SRNBRK.

\bibliography{Bib}

\end{document}